\documentclass[12pt]{article}
\usepackage[cp1250]{inputenc}
\usepackage[verbose,a4paper,tmargin=0.5in,lmargin=1in,rmargin=1in]{geometry}
\usepackage[pdftex]{graphicx}
\usepackage{amsfonts}
\usepackage{indentfirst}
\usepackage{latexsym}
\usepackage{amsmath}
\usepackage{amssymb}
\usepackage{psfrag}
\usepackage{eufrak}
\usepackage[mathscr]{eucal} 

\title{Classification of the Killing Vectors in Nonexpanding $\mathcal{HH}$-Spaces with $\Lambda$.}

\author{$\textrm{Adam Chudecki}^{*} \dag$}

\begin{document}

\maketitle

$*$ Center of Mathematics and Physics  

$\ \ $ Technical University of Łódź           

$\ \ $ Al. Politechniki 11, 90-924 Łódź, Poland
\newline

$\dag$ E-mail: adam.chudecki@p.lodz.pl
\newline
\newline
\textbf{Abstract.} 
Conformal Killing equations and their integrability conditions for nonexpanding hyperheavenly spaces with $\Lambda$ are studied. Reduction of ten Killing equations to one \textsl{master equation} is presented. Classification of homothetic and isometric Killing vectors in nonexpanding hyperheavenly spaces with $\Lambda$ and homothetic Killing vectors in heavenly spaces is given. Some nonexpanding complex metrics of types $[\textrm{III,N}] \otimes [\textrm{N}]$ are found. A simple example of Lorentzian real slice of the type $[\textrm{N}] \otimes [\textrm{N}]$ is explicitly given.

\section{Introduction}

This is a third part of our work devoted to the conformal, homothetic and isometric Killing symmetries in hyperheavenly spaces. In the first part \cite{biblio_51} the nonexpanding hyperheavenly spaces have been considered and in fact, that part is the generalization of the work by Finley and Plebański \cite{biblio_27}. The Killing symmetries in expanding hyperheavenly spaces have been presented in our second paper \cite{biblio_54} which can be considered as a development of the ideas and results given by Sonnleitner and Finley \cite{biblio_55}. Namely, in our work \cite{biblio_54} the nonzero cosmological constant $\Lambda$ has been included and some more compact classification of the Killing vectors has been done. 

The current paper is a continuation of \cite{biblio_51} and a generalization of the outstanding work by Finley and Plebański \cite{biblio_16}. Here we present classification of the homothetic and isometric Killing vectors in nonexpanding hyperheavenly spaces. We show how to use the general results to classify the homothetic Killing vectors in heavenly spaces. Such a classification together with different types of the reduced heavenly equation (admitting at least one homothetic Killing vector) has not been considered by Plebański and co-workers (isometric Killing vectors in heavenly spaces has been found in \cite{biblio_16}). Moreover, we have been able to obtain some Lorentzian real slices. This last result follows from the previous considerations in such a natural manner that it seems to be very promising in the solution of a long studying fundamental problem: how to find a Lorentzian spacetime from a given complex spacetime.

The generalization of the heavenly spaces, i.e. hyperheavenly spaces were discovered by Plebański and Robinson in 1976 \cite{biblio_18} - \cite{biblio_19}. In the late seventies and early nineties many papers devoted to this spaces were published. Structure of the heavenly \cite{biblio_11} - \cite{biblio_15} and hyperheavenly \cite{biblio_20} - \cite{biblio_23} spaces, especially in the spinorial formalism has been investigated with details. General properties of the Killing spinors \cite{biblio_26} and two-sided algebraically degenerated hyperheavenly spaces \cite{biblio_24} have complemented the image of the complex general relativity. 

The most important question: how to use the complex solutions of the Einstein field equation to obtain solutions of the real spacetimes with Lorentzian signature still remains without answer. The theoretical properties of such a slices have been presented in \cite{biblio_17}. Different approach \cite{biblio_56} used the connection 1-form to assure the existence of the real Lorentzian slices. The difficulties in obtaining the real solutions with the Lorentzian signature were the main reason, why the complex relativity theory seemed to be of a little interest in nineties. 

Recently some compendium of the hyperheavenly spaces has been presented \cite{biblio_29}. Results from \cite{biblio_24} was used to obtain the special form of the hyperheavenly equation determining all $[\textrm{N}] \otimes [\textrm{N}]$-type spaces with nonzero twist \cite{biblio_57}. Although obtaining the real spacetimes with Lorentzian signature is a very difficult task, the real spacetimes with ultrahyperbolic signature $(++--)$ can be obtained very easily. That is why hyperheavenly machinery appeared to be useful tool in Walker and Osserman spaces \cite{biblio_32}, \cite{biblio_33}. 
 
The problem of Killing symmetries given by the set of equations $\nabla_{(a} K_{b)} = \chi \, g_{ab}$ in nonexpanding hyperheavenly spaces was presented in \cite{biblio_27} and then in \cite{biblio_51}. In \cite{biblio_21} the nonzero cosmological constant $\Lambda$ and nonconstant conformal factor $\chi$ was introduced. In the present paper we show how to classify the Killing vectors in nonexpanding hyperheavenly spaces. Moreover, some new, more compact way of reduction of the Killing equations to one master equation is found.

In the present paper we use the following terminology. Equation $\nabla_{(a} K_{b)} = \chi \, g_{ab}$ is called the conformal Killing equation and its solution $K$ the conformal Killing vector. If $\chi = \chi_{0} = \textrm{const}$ then we deal with the homothetic Killing equation and the homothetic Killing vector. Finally, if $\chi=0$ we have the isometric Killing equation and the isometric Killing vector, respectively.

Our paper is organized as follows. In section 2 the general structure of nonexpanding hyperheavenly space is presented. We recall the form of the hyperheavenly equation, connection forms and curvature. In section 3 some final results, especially the form of the master equation and the transformation formulas are presented. Section 4 is devoted to the classification of the isometric and homothetic Killing vectors. Some examples of the nonexpanding hyperheavenly spaces admitting isometric, null Killing vector field are given. One of them appear to be a complex extension of the pp-wave. Finally, the Lorentzian real slice of the hyperheavenly space of the type $[\textrm{N}] \otimes [\textrm{N}]$ is found. In section 5 we show, how to use the results from the section 3 to describe the homothetic Killing symmetries in heavenly spaces. Moreover, the classification of such symmetries is explicitly given. Detailed reduction of the Killing problem for the expanding hyperheavenly spaces are done in section 6. Concluding remarks end our paper. 

Our work give some new hope of positive results in so called Plebański program, namely: how one can generate real Lorentzian metrics from holomorphic ones. Until now, that step seemed to be very complicated. The simplicity of obtaining the real pp-wave from the complex solution shows, that in appropriate coordinate frame Lorentzian real slice could be found very fast. Application of this statement to the expanding hyperheavenly spaces, especially of the type $[\textrm{N}] \otimes [\textrm{N}]$ will become one of our main research program.

\section{Hyperheavenly spaces.}

\subsection{General structure of hyperheavenly spaces.}

\textsl{$\mathcal{HH}$-space with cosmological constant} is a 4 - dimensional complex analytic differential manifold $\mathcal{M}$ endowed with a holomorphic Riemannian metric $ds^{2}$ satisfying the vacuum Einstein equations with cosmological constant and such that the self - dual or anti - self - dual part of the Weyl tensor is algebraically degenerate \cite{biblio_18, biblio_19, biblio_20}. These kind of spaces admits a congruence of totally null, self-dual (or anti-self-dual, respectively) surfaces, called \textsl{null strings} \cite{biblio_28}. In this paper we deal with self-dual null strings. Coordinate system can be always chosen such that $\Gamma_{422}=\Gamma_{424}=0$ \cite{biblio_19}, the surface element of null string is given by $e^{1} \wedge e^{3}$, and null tetrad $(e^{1},e^{2},e^{3},e^{4})$ in spinorial notation can be chosen as
\begin{eqnarray}
\label{tetrada_spinorowa}
e_{\dot{A}} := \phi^{-2} \, dq_{\dot{A}} &=& 
\left[ \begin{array}{c}  e^{3}  \\ e^{1}  \end{array} \right]  \ \ = \ \ 
-\frac{1}{\sqrt{2}} \, g^{2}_{\ \, \dot{A}}
\\ \nonumber
E^{\dot{A}} := -dp^{\dot{A}} + Q^{\dot{A} \dot{B}} \, dq_{\dot{B}} &=&
\left[ \begin{array}{c}  e^{4}  \\ e^{2}  \end{array} \right]  \ \ = \ \ 
\frac{1}{\sqrt{2}} \, g^{1 \dot{A}}
\end{eqnarray}
where
\begin{equation}
(g^{A\dot{B}}) := \sqrt{2}
\left[\begin{array}{cc}
e^4 & e^2 \\
e^1 & -e^3
\end{array}\right] 
\end{equation}
$\phi$ and $Q^{\dot{A} \dot{B}}$ are holomorphic functions. Spinorial coordinates $p^{\dot{A}}$ are coordinates on null strings given by $q_{\dot{A}}=\textrm{const}$.
Define the following operators
\begin{eqnarray}
&&\partial_{\dot{A}} := \frac{\partial}{\partial p^{\dot{A}}}
\ \ \ \ \ \ \ \ \ \ \ \ \ \ \ \ \ \ \ \ \ \ \ \ 
\partial^{\dot{A}} := \frac{\partial}{\partial p_{\dot{A}}}
\\ \nonumber
&&\eth^{\dot{A}} := \phi^{2} \left( \frac{\partial}{\partial q_{\dot{A}}} + Q^{\dot{A} \dot{B}} \partial_{\dot{B}} \right)
\ \ \ \ \ \ \ \ 
\eth_{\dot{A}} := \phi^{2} \left( \frac{\partial}{\partial q^{\dot{A}}} - Q_{\dot{A}}^{\ \ \dot{B}} \partial_{\dot{B}} \right)
\end{eqnarray}
They form the basis dual to $(e_{\dot{A}}, E^{\dot{B}})$
\begin{equation}
\label{baza_dualna}
-\partial_{\dot{A}} = 
\left[ \begin{array}{c}  \partial_{4}  \\ \partial_{2}  \end{array} \right] 
 \ \ \ \ \ \ \ \ 
\eth^{\dot{A}} = 
\left[ \begin{array}{c}  \partial_{3}  \\ \partial_{1}  \end{array} \right] 
\end{equation}
Spinorial indices are to be manipulated according to the rules $q_{\dot{A}} = \ \in_{\dot{A} \dot{B}} q^{\dot{B}}$, $q^{\dot{A}} = q_{\dot{B}} \in^{\dot{B} \dot{A}}$. Then, the rules to raise and lower spinor indices in the case of objects from tangent space read $\partial^{\dot{A}} = \partial_{\dot{B}} \in^{\dot{A} \dot{B}}$, $\partial_{\dot{A}} = \in_{\dot{B} \dot{A}} \partial^{\dot{B}}$, $\eth^{\dot{A}} = \eth_{\dot{B}} \in^{\dot{A} \dot{B}}$, $\eth_{\dot{A}} = \in_{\dot{B} \dot{A}} \eth^{\dot{B}}$. As usual, $\in_{AB}$ and $\in_{\dot{A}\dot{B}}$ are spinorial Levi-Civita symbols
\begin{eqnarray}
&&( \in_{AB} ) := \left[ \begin{array}{cc}
                            0 & 1   \\
                           -1 & 0  
                            \end{array} \right] =: ( \in^{AB} )
\ \ \ , \ \ \ 
( \in_{\dot{A}\dot{B}} ) := \left[ \begin{array}{cc}
                            0 & 1   \\
                           -1 & 0  
                            \end{array} \right] =: ( \in^{\dot{A}\dot{B}} ) 
\\ \nonumber
&& \in_{AC} \in^{AB} = \delta^{B}_{C} \ \ \ , \ \ \ \in_{\dot{A}\dot{C}} \in^{\dot{A}\dot{B}} = \delta^{\dot{B}}_{\dot{C}} \ \ \ , \ \ \ 
(\delta^{A}_{C})= (\delta^{\dot{B}}_{\dot{C}})= \left[ \begin{array}{cc}
                            1 & 0   \\
                            0 & 1  
                            \end{array} \right]
\end{eqnarray}
The metric is given by
\begin{equation}
\label{metryka_ogolna}
ds^{2} =2 \, e^{1} \underset{s}{\otimes} e^{2} + 2 \, e^{3} \underset{s}{\otimes} e^{4}
= -\frac{1}{2} \, g_{A\dot{B}} \underset{s}{\otimes} g^{A\dot{B}}
= 2\phi^{-2} \, (- dp^{\dot{A}} \underset{s}{\otimes} dq_{\dot{A}} + 
           Q^{\dot{A} \dot{B}} \, dq_{\dot{A}} \underset{s}{\otimes} dq_{\dot{B}})
\end{equation}
The expansion 1-form which characterises congruence of self-dual null strings is defined by \cite{biblio_20}
\begin{equation}
\label{forma_ekspansji}
\theta := \theta_{a} e^{a} = \phi^{4} (\Gamma_{423} e^{3} + \Gamma_{421} e^{1}) = \phi \, \frac{\partial \phi}{\partial p_{\dot{A}}} \, dq_{\dot{A}}
\end{equation}
Consequently, we can consider two cases
\begin{itemize}
\item $\frac{\partial \phi}{\partial p_{\dot{A}}}=0 \rightarrow \theta=0$ and such a space is called \textsl{nonexpanding $\mathcal{HH}$-space} (in this case one can put $\phi=1$)
\item $\frac{\partial \phi}{\partial p_{\dot{A}}}\ne0 \rightarrow \theta \ne 0$; such a space is called  \textsl{expanding $\mathcal{HH}$-space}
\end{itemize}
In the present paper we deal with nonexpanding case.

\subsection{Nonexpanding hyperheavenly spaces.}

By reducing Einstein equations \cite{biblio_19, biblio_20} one gets $\phi=1$ and
\begin{subequations}
\begin{eqnarray}
Q^{\dot{A} \dot{B}} &=& - \Theta_{p_{\dot{A}}p_{\dot{B}}} + \frac{2}{3} F^{(\dot{A}}p^{\dot{B})} + \frac{1}{3} \Lambda \, p^{\dot{A}} p^{\dot{B}}
\\
\label{expanding_QAB}
&&=\partial^{(\dot{A}} \Omega^{\dot{B})} + \frac{1}{3} \Lambda \, p^{\dot{A}} p^{\dot{B}}
\end{eqnarray}
\end{subequations}
where
\begin{equation}
\label{postac_Omegi}
\Omega^{\dot{B}} 
= - \Theta_{p_{\dot{B}}} - \frac{2}{3} F_{\dot{A}} p^{\dot{A}} p^{\dot{B}}
\end{equation}
where $F^{\dot{A}}$ is an arbitrary function of $q^{\dot{A}}$ only, and $\Theta$ is an arbitrary function of all variables called \textsl{the key function}.
Einstein equations can be reduced to one equation called the \textsl{nonexpanding hyperheavenly equation with $\Lambda$}
\begin{equation}
\mathcal{T} + F^{\dot{A}} \Theta_{p^{\dot{A}}} + \frac{1}{6} (F^{\dot{A}} p_{\dot{A}})^{2} + \Theta_{p_{\dot{A}} q^{\dot{A}}} + \frac{1}{6} \frac{\partial F_{\dot{A}}}{\partial q^{\dot{B}}} p^{\dot{A}} p^{\dot{B}} + \Lambda \big( p^{\dot{A}} \Theta_{p^{\dot{A}}} - \Theta \big) = N_{\dot{A}} p^{\dot{A}} + \gamma
\end{equation}
where
\begin{equation}
\mathcal{T} := \frac{1}{2} Q^{\dot{A} \dot{B}} Q_{\dot{A} \dot{B}} 
\end{equation}
Inserting the explicit form of $\mathcal{T}$
\begin{eqnarray}
\label{nonexpanding_hyperheavenly_equation}
\frac{1}{2} \, \Theta_{p_{\dot{A}} p_{\dot{B}}} \Theta_{p^{\dot{A}} p^{\dot{B}}}
+ \Theta_{p_{\dot{A}} q^{\dot{A}}} 
+F^{\dot{A}} \Big( \Theta_{p^{\dot{A}}} - \frac{2}{3} \, p^{\dot{B}} \, \Theta_{p^{\dot{A}} p^{\dot{B}}}  \Big) + \frac{1}{18} \, (F^{\dot{A}} p_{\dot{A}})^{2}&&
\\ \nonumber
+ \frac{1}{6} \, \frac{\partial F_{\dot{A}}}{\partial q^{\dot{B}}} \, p^{\dot{A}} p^{\dot{B}} 
+\Lambda \Big( p^{\dot{A}} \Theta_{p^{\dot{A}}} - \Theta - \frac{1}{3} \, p^{\dot{A}} p^{\dot{B}} \Theta_{p^{\dot{A}} p^{\dot{B}}}   \Big)
&=& N_{\dot{A}} \, p^{\dot{A}} +\gamma \ \ \ \ \ \ \ \ \ 
\end{eqnarray}
where $N^{\dot{A}}$ and $\gamma$ are arbitrary functions of $q^{\dot{C}}$ only (constant on each null string), and $\Theta_{p^{\dot{A}}} \equiv \frac{\partial \Theta}{\partial p^{\dot{A}}}$,  $\Theta_{q_{\dot{A}}} \equiv \frac{\partial \Theta}{\partial q_{\dot{A}}}, etc.$.
Explicit formulas for the connection 1-forms in spinorial formalism read
\begin{eqnarray}
\label{expanding_koneksja}
\mathbf{\Gamma}_{11}  &=& 0                   
\\ \nonumber
\mathbf{\Gamma}_{12} &=& 
-\frac{1}{2}  \, \partial^{\dot{A}}  Q_{\dot{A} \dot{B}} \, e^{\dot{B}}
\\ \nonumber
\mathbf{\Gamma}_{22} &=& - \eth^{\dot{A}} Q_{\dot{A}\dot{B}} \, e^{\dot{B}} 
\\ \nonumber
\mathbf{\Gamma}_{\dot{A} \dot{B}} &=& 
 \partial_{(\dot{A}} Q_{\dot{B})\dot{C}} \, e^{\dot{C}} 
\end{eqnarray}
where (as a consequence of hyperheavenly equation, especially useful in many calculations)
\begin{equation}
\eth_{\dot{A}} Q^{\dot{A}\dot{B}} = N^{\dot{B}} - \frac{1}{2} \, p^{\dot{B}} \,  C^{(2)}
\end{equation}
Decomposing the connection 1-forms according to
\begin{equation}
\label{rozklad_koneksji}
\mathbf{\Gamma}_{AB} = -\frac{1}{2} \, \mathbf{\Gamma}_{AB C \dot{D}} \, g^{C \dot{D}}
\ \ \ \ \ \ \ \ \  
\mathbf{\Gamma}_{\dot{A} \dot{B}} = -\frac{1}{2} \, \mathbf{\Gamma}_{\dot{A} \dot{B} C \dot{D}} \, g^{C \dot{D}}
\end{equation}
we get
\begin{eqnarray}
&&\mathbf{\Gamma}_{111 \dot{D}} =0  \ \ \ \ \ \ \ \ \ \ \ \ \ \ \ \ \  \ \ \ \ \ \ \ \ \ \ \ \ \ 
\mathbf{\Gamma}_{112 \dot{D}} = 0
\\ \nonumber
&&\mathbf{\Gamma}_{121 \dot{D}} = 0 \ \ \ \ \ \ \ \ \ \ \ \ \ \ \ \ \  \ \ \ \ \ \ \ \ \ \ \ \ \ 
  \mathbf{\Gamma}_{122 \dot{D}} = - \frac{1}{\sqrt{2}} \,  \partial^{\dot{A}}  Q_{\dot{A}\dot{D}}
\\ \nonumber
&& \mathbf{\Gamma}_{221 \dot{D}} = 0
\ \ \ \ \ \ \ \ \ \ \ \ \ \ \ \ \ \ \ \ \ \ \ \ \ \ \ \ \  \ 
\mathbf{\Gamma}_{222 \dot{D}} = -\sqrt{2}  \, \eth^{\dot{A}} Q_{\dot{A}\dot{D}}
\\ \nonumber
&&\mathbf{\Gamma}_{\dot{A} \dot{B} 1 \dot{D}} = 0
\ \ \ \ \ \  \ \ \ \ \ \ \ \ \ \ \ \ \  \ \ \ \ \ \ \ \ \ \ 
\mathbf{\Gamma}_{\dot{A} \dot{B} 2 \dot{D}} =  \sqrt{2} \, \partial_{(\dot{A}} Q_{\dot{B})\dot{D}}  
\end{eqnarray}
The conformal curvature is given by
\begin{eqnarray}
\label{nonexpanding_krzywizna}
&&C^{(5)}  = 0 = C^{(4)} \ \ \ \ \ \ \ \ \ \ \ \  \ \ \ \ \ \
C^{(3)}  = -\frac{2}{3} \, \Lambda \ \ \ \ \ \ \ \ \ \ \ \  \ \ \ \ \ \
C^{(2)}  =  -\frac{\partial F^{\dot{A}}}{\partial q^{\dot{A}}}  \ \ \ \ \ \ \  \ \ \ \ \ \
\\ \nonumber
&&C^{(1)} =
 2  F^{\dot{A}} N_{\dot{A}} - 2 \, \frac{\partial N_{\dot{A}}}{\partial q_{\dot{A}}}   + Z_{\dot{A}} p^{\dot{A}}
\\ \nonumber
&&C_{\dot{A}\dot{B}\dot{C}\dot{D}} = 
\Theta_{p^{\dot{A}}p^{\dot{B}}p^{\dot{C}}p^{\dot{D}}}
\end{eqnarray}
where
\begin{equation}
\label{definicja_Z}
Z_{\dot{A}} := F_{\dot{A}} C^{(2)} + \frac{\partial C^{(2)}}{\partial q^{\dot{A}}} + 2 \Lambda \, N_{\dot{A}}
\end{equation}

\subsection{Gauge freedom.}
\label{subsection_gauge}

The problem of coordinate gauge freedom in nonexpanding hyperheavenly spaces was considered in details in \cite{biblio_51, biblio_19, biblio_20, biblio_33}.  The form of metric (\ref{metryka_ogolna}) admits the coordinate gauge freedom
\begin{eqnarray}
\label{gaug_wspolrzednych}
q'_{\dot{A}} &=& q'_{\dot{A}} (q_{\dot{B}})
\\ \nonumber
p'^{\dot{A}} &=&  \frac{\partial q_{\dot{B}}}{\partial q'_{\dot{A}}} \, p^{\dot{B}} + \sigma^{\dot{A}} 
\ \ \ \ , \ \ \ \ 
 \sigma^{\dot{A}}=\sigma^{\dot{A}} (q_{\dot{B}})
\end{eqnarray}
where $\sigma^{\dot{A}}$ are arbitrary functions of $q^{\dot{A}}$ only. Denote
\begin{eqnarray}
D_{\dot{A}}^{\ \ \dot{B}} 
&=& \frac{\partial q'_{\dot{A}}}{\partial q_{\dot{B}}} 
= \Delta \, \frac{\partial q^{\dot{B}}}{\partial q'^{\dot{A}}}
=  \Delta \, \frac{\partial p'_{\dot{A}}}{\partial p_{\dot{B}}}
=  \frac{\partial p^{\dot{B}}}{\partial p'^{\dot{A}}}
\\ \nonumber
D^{-1 \ \ \dot{B}}_{\ \ \; \dot{A}} 
&=& \frac{\partial q_{\dot{A}}}{\partial q'_{\dot{B}}}
= \Delta^{-1} \, \frac{\partial q'^{\dot{B}}}{\partial q^{\dot{A}}}
=  \frac{\partial p'^{\dot{B}}}{\partial p^{\dot{A}}}
=  \Delta^{-1} \, \frac{\partial p_{\dot{A}}}{\partial p'_{\dot{B}}}
\end{eqnarray}
where $\Delta$ stands for the determinant of the matrix $D_{\dot{A}}^{\ \ \dot{B}} $
\begin{equation}
\label{definicja_wyznacznika_macierzy_przejscia}
\Delta := \det 
\left( \frac{\partial q'_{\dot{A}}}{\partial q_{\dot{B}}} \right)
 = \frac{1}{2} \, D_{\dot{A}\dot{B}} D^{\dot{A}\dot{B}}
\end{equation}
The functions $Q^{\dot{A}\dot{B}}$ transform under (\ref{gaug_wspolrzednych}) as follows
\begin{equation}
Q'^{\dot{A}\dot{B}} 
=D^{-1 \ \ \dot{A}}_{\ \ \; \dot{R}} \, D^{-1 \ \ \dot{B}}_{\ \ \; \dot{S}}
Q^{\dot{R} \dot{S}} + D^{-1 \ \ ( \dot{A}}_{\ \ \; \dot{R}} \, 
\frac{\partial p'^{\dot{B})}}{\partial q_{\dot{R}}} 
\end{equation}
It is easy to see that the transformation (\ref{gaug_wspolrzednych}) is equivalent to the spinorial transformation of the tetrad with 
\begin{eqnarray}
\label{transformacja_spinorowa}
L^{A}_{\ \; B} &=& \left[ \begin{array}{cc}
       \Delta^{-\frac{1}{2}} & h  \Delta^{\frac{1}{2}}   \\
       0 &  \Delta^{\frac{1}{2}}  
       \end{array} \right]
\\ \nonumber
M^{\dot{A}}_{\ \; \dot{B}} &=& \Delta^{\frac{1}{2}} \, D^{-1 \ \ \dot{A}}_{\ \ \; \dot{B}}
\end{eqnarray}
where
\begin{equation}
2h := -  D^{-1 \ \ \dot{R}}_{\ \ \; \dot{S}} \, 
\frac{\partial p^{\dot{S}}}{\partial q'^{\dot{R}}} 
= \frac{1}{\Delta} \, D_{\dot{S}}^{\ \ \dot{R}} \, 
\frac{\partial p'^{\dot{S}}}{\partial q^{\dot{R}}} 
= \frac{\partial \sigma^{\dot{R}}}{\partial q'^{\dot{R}}} 
\end{equation}
[If $\Psi^{A...\dot{B}...}$ is a spinor quantity, it transforms according to formula 
\begin{equation}
\nonumber
\Psi'^{A...\dot{B}...}=L^{A}_{\ \; R}...M^{\dot{B}}_{\ \; \dot{S}}...\Psi^{R...\dot{S}...} \ \ \ \ ]
\end{equation}
By demanding, that the connection forms, curvature coefficient and the hyperheavenly equation has the same form in new coordinate frame, one can get the transformation formulas for the structural functions. Straightforward, but somewhat tedious calculations lead to
\begin{subequations}
\begin{eqnarray}
\label{transformacja_F}
F'^{\dot{A}} &=& D^{-1 \ \ \dot{A}}_{\ \ \; \dot{R}} \, F^{\dot{R}} 
- \frac{\partial \ln \Delta }{\partial q'_{\dot{A}}} -\Lambda \, \sigma^{\dot{A}}
\\ 
\label{transformacja_N}
N'^{\dot{A}} &=& \Delta^{-1} \, D^{-1 \ \ \dot{A}}_{\ \ \; \dot{R}} \, N^{\dot{R}}
+ \frac{\partial h}{\partial q'_{\dot{A}}}
- h F'^{\dot{A}} 
- \frac{1}{2} \, \frac{\partial F'^{\dot{S}}}{\partial q'^{\dot{S}}} \, \sigma^{\dot{A}}
-\Lambda \, h \sigma^{\dot{A}}
\\ 
\label{transformacja_gammy_prostrzej}
\Delta^2 \gamma' &=& \gamma + \frac{\partial H^{\dot{A}}}{\partial q^{\dot{A}}} 
+ F_{\dot{A}} H^{\dot{A}}
+\frac{1}{6} \Delta^{2} \, \sigma^{\dot{A}} \sigma^{\dot{B}} \, \frac{\partial F'_{\dot{A}}}{\partial q'^{\dot{B}}}
+\frac{1}{18} \Delta^{2} \, (F'^{\dot{A}} \sigma_{\dot{A}})^{2} 
\ \ \ \ \ \ \ \ \ \ 
\\ \nonumber
&&-\frac{1}{2} N_{\dot{A} \dot{B}} N^{\dot{A} \dot{B}} 
- \Delta^{2} N'_{\dot{A}} \sigma^{\dot{A}} - \Lambda \, M
\\
\label{transformacja_fun_klucz_dla_nonexpan}
\Delta^{2} \Theta' &=& \Theta + \frac{1}{6} \, L^{\dot{A}\dot{B}\dot{C}} p_{\dot{A}} p_{\dot{B}} p_{\dot{C}} 
+ \frac{1}{2} \, N^{\dot{A}\dot{B}} p_{\dot{A}} p_{\dot{B}} 
+ H^{\dot{A}}p_{\dot{A}} + M
\end{eqnarray}
\end{subequations}
where
\begin{eqnarray}
\label{definicje_funkcyj_w_gaugu_dla_nonexpan}
L^{\dot{A} \dot{B}}_{\ \ \ \, \dot{C}} &:=& 
D_{\dot{X}}^{\ \ (\dot{A}} \frac{\partial }{\partial q_{\dot{B})}}
\Big(D^{-1 \ \ \dot{X}}_{\ \ \; \dot{C}} \Big)
+\frac{2}{3} \, \delta^{( \dot{A}}_{\dot{C}} \, 
  \frac{\partial \ln \Delta}{\partial q_{\dot{B})}}
\\ \nonumber
&=& D_{\dot{X}}^{\ \ \dot{A}} D_{\dot{Y}}^{\ \ \dot{B}} 
    \frac{\partial^{2} q_{\dot{C}}}{\partial q'_{\dot{X}} \partial q'_{\dot{Y}}} 
+\frac{2}{3} \, \delta^{( \dot{A}}_{\dot{C}} \, 
  \frac{\partial \ln \Delta}{\partial q_{\dot{B})}}
\\ \nonumber
N^{\dot{A} \dot{B}} &:=& D_{\dot{X}}^{\ \ \dot{A}} D_{\dot{Y}}^{\ \ \dot{B}} 
\bigg(  
\frac{2}{3} \, F'^{( \dot{X}} \sigma^{\dot{Y})} 
- \frac{\partial \sigma^{(\dot{X}}}{\partial q'_{\dot{Y})}}
+ \frac{1}{3} \, \Lambda \, \sigma^{\dot{X}} \sigma^{\dot{Y}}
 \bigg)
\end{eqnarray}
$H^{\dot{A}}=H^{\dot{A}}(q_{\dot{B}})$ and $M=M(q_{\dot{B}})$ are arbitrary functions of $q_{\dot{B}}$ only.
\newline
(Formulas (\ref{transformacja_F}), (\ref{transformacja_N}), (\ref{transformacja_gammy_prostrzej}), (\ref{transformacja_fun_klucz_dla_nonexpan}) and (\ref{definicje_funkcyj_w_gaugu_dla_nonexpan}) generalize the respective results of \cite{biblio_20, biblio_22} on the case when $\Lambda \ne 0$ and they were presented in \cite{biblio_29} and then in \cite{biblio_33}.)
\newline

\setlength\arraycolsep{2pt}
\setcounter{equation}{0}

\section{Conformal Killing symmetries.}
\label{sekcja_Killing}

\subsection{Conformal Killing equations and their integrability conditions in spinorial formalism.}

Define the spin-tensor $g^{aA \dot{B}}$ by the relation $g^{A \dot{B}} = g_{a}^{\ A \dot{B}} \, e^{a}$. Hence, $-\frac{1}{2} g^{aA\dot{B}}g_{bA\dot{B}} = \delta^{a}_{b}$ and $-\frac{1}{2}g^{aA\dot{B}}g_{aC\dot{D}} = \delta^{A}_{C} \delta^{\dot{B}}_{\dot{D}}$. The operators $\partial^{A\dot{B}}$ and $\nabla^{A\dot{B}}$ are spinorial equivalences of operators $\partial^{a}$ and $\nabla^{a}$, respectively, given by
\begin{equation}
\partial^{A\dot{B}} = g_{a}^{\  A \dot{B}} \partial^{a}  \ \ \ \ \ \ \ \ \ \ \ \ \ \ 
\nabla^{A\dot{B}} = g_{a}^{\  A \dot{B}} \nabla^{a} 
\end{equation}
In the basis (\ref{baza_dualna})
\begin{equation}
\label{rozklad_operatora}
\partial^{A\dot{B}} = \sqrt{2} \, (\delta^{A}_{1} \, \eth^{\dot{B}} - \delta^{A}_{2} \, \partial^{\dot{B}}) = \sqrt{2} \, [ \eth^{\dot{B}}, -\partial^{\dot{B}}]
\end{equation}
Killing vector has the form
\begin{equation}
K = K^{a} \, \partial_{a} = -\frac{1}{2} K_{A\dot{B}} \partial^{A\dot{B}} = k_{\dot{B}} \eth^{\dot{B}} - h_{\dot{B}} \partial^{\dot{B}}
\end{equation}
where we use the decomposition
\begin{equation}
\label{rozklad_wektora_Killinga}
K_{A\dot{B}} = - \sqrt{2} \, (\delta^{1}_{A} \,k_{\dot{B}} + \delta^{2}_{A} \,h_{\dot{B}} ) \,  
= - \sqrt{2} \,  [k_{\dot{B}},h_{\dot{B}}]
\end{equation}
Components of the Killing vector, $K^{a}$ and $K_{A \dot{B}}$ are related by the condition
\begin{equation}
K^{a} = - \frac{1}{2} g^{a A \dot{B}} \, K_{A \dot{B}} \ \ \leftrightarrow \ \ 
K_{A \dot{B}} = g_{a  A \dot{B}} \, K^{a}
\end{equation}
Conformal Killing equations with a conformal factor $\chi$ read
\begin{equation}
\label{podstawowe_nasze_rownania_Killinga}
\nabla_{(a} K_{b)} = \chi \, g_{ab}
\end{equation}
In spinorial form
\begin{equation}
\nabla_{A}^{\ \; \dot{B}} K_{C}^{\ \dot{D}} + \nabla_{C}^{\ \; \dot{D}} K_{A}^{\ \dot{B}} = -4 \chi \in_{AC} \in^{\dot{B}\dot{D}}
\end{equation}
what is equivalent to the following equations
\begin{subequations}
\begin{eqnarray}
\label{rozklad_rownania_Killinga}
E_{AC}^{\ \ \ \dot{B}\dot{D}} &\equiv& \nabla_{(A}^{\ \; \dot{(B}} K_{C)}^{\ \; \dot{D)}} = 0 
\\
\label{definicja_chi}
E &\equiv& \nabla^{N\dot{N}} K_{N\dot{N}} +8 \chi = 0
\end{eqnarray}
\end{subequations}
From (\ref{rozklad_rownania_Killinga}) and (\ref{definicja_chi}) it follows that
\begin{equation}
\nabla_{A}^{\ \; \dot{B}} K_{C}^{\ \; \dot{D}} = l_{AC} \in^{\dot{B}\dot{D}} + l^{\dot{B}\dot{D}} \in_{AC} - 2 \chi  \in_{AC} \in^{\dot{B}\dot{D}}
\end{equation}
with
\begin{equation}
l_{AC} := \frac{1}{2} \nabla_{(A}^{\ \ \dot{N}} K_{C)\dot{N}} \ \ \ \ \ \ \ \ \ \ \ \ 
l^{\dot{B}\dot{D}} := \frac{1}{2} \nabla^{N(\dot{B}} K_{N}^{\ \; \dot{D})}
\end{equation}
In \cite{biblio_26} the integrability conditions of (\ref{rozklad_rownania_Killinga}) and (\ref{definicja_chi}) have been found. For the Einstein space ($C_{AB\dot{C}\dot{D}}=0$, $R=-4 \Lambda$) these conditions consist of the following equations
\begin{subequations}
\begin{eqnarray}
\label{integrability_c_L}
L_{RST}^{\ \ \ \ \; \dot{A}} &\equiv& \nabla_{R}^{\ \; \dot{A}} l_{ST} + 2C^{N}_{\ RST} K_{N}^{\ \; \dot{A}} + \frac{2}{3} \Lambda \in_{R(S}K_{T)}^{\ \ \dot{A}} + 2 \in_{R(S} \nabla_{T)}^{\ \ \dot{A}} \chi =0 \ \ \ \ \ \ \ \ 
\\
\label{integrability_c_M}
M_{ABCD} &\equiv& K_{N\dot{N}} \nabla^{N\dot{N}}C_{ABCD} + 4C^{N}_{\ \; (ABC} l_{D)N} - 4 \chi C_{ABCD} = 0
\\
\label{integrability_c_N}
N_{AB}^{\ \ \ \dot{A}\dot{B}} &\equiv& \nabla_{A}^{\ \; \dot{A}} \nabla_{B}^{\ \; \dot{B}} \chi - \frac{2}{3} \Lambda \chi \in_{AB} \in^{\dot{A}\dot{B}} =0
\\
\label{integrability_c_R}
R_{ABC}^{\ \ \ \ \ \dot{A}} &\equiv& C^{N}_{\ \; ABC} \nabla_{N}^{\ \; \dot{A}} \chi =0
\end{eqnarray}
\end{subequations}
for undotted $l_{AB}$ and $C_{ABCD}$ and the respective equations $L_{\dot{R}\dot{S}\dot{T}}^{\ \ \ \ \; A}$, $M_{\dot{A}\dot{B}\dot{C}\dot{D}}$ and $R_{\dot{A}\dot{B}\dot{C}}^{\ \ \ \ \ A}$ for dotted $l_{\dot{A}\dot{B}}$ and $C_{\dot{A}\dot{B}\dot{C}\dot{D}}$.

\subsection{Final results.}

Explicit form of the conformal Killing equations and its integrability conditions, together with the way of reduction the Killing equations to the one master equation are presented in section \ref{redukcja_rownan_killinga}. Finally, we obtain:
\newline
Spinors $l_{AB}$
\begin{subequations}
\begin{eqnarray}
l_{11} &=& 0
\\
l_{12} &=& \delta_{\dot{N}} (F^{\dot{N}} + \Lambda p^{\dot{N}}) - 2 \chi + \frac{\partial \delta^{\dot{N}}}{\partial q^{\dot{N}}}
\\
l_{22} &=& -C^{(2)} \, \delta_{\dot{N}} p^{\dot{N}} -2 \, \frac{\partial \chi}{\partial q^{\dot{N}}} p^{\dot{N}} + 2 \delta_{\dot{N}} N^{\dot{N}} - \frac{\partial \epsilon^{\dot{N}}}{\partial q^{\dot{N}}}
\end{eqnarray}
\end{subequations}
Components of the Killing vector read
\begin{eqnarray}
\label{componentss_off_the_Killing_vector}
k^{\dot{A}} &=& \delta^{\dot{A}} (q^{\dot{M}})
\\
h^{\dot{A}} &=& \delta_{\dot{S}} Q^{\dot{S}\dot{A}} - 2\chi \, p^{\dot{A}} - \frac{\partial \delta^{\dot{N}}}{\partial q_{\dot{A}}} \, p_{\dot{N}} - \epsilon^{\dot{A}}
\end{eqnarray}
what gives the form of the Killing vector
\begin{equation}
K = \delta^{\dot{B}} \frac{\partial}{\partial q^{\dot{B}}} + \bigg(2\chi \, p^{\dot{B}} +\frac{\partial \delta^{\dot{M}}}{\partial q_{\dot{B}}} p_{\dot{M}} + \epsilon^{\dot{B}} \bigg) \frac{\partial}{\partial p^{\dot{B}}} 
\end{equation}
Ten conformal Killing equations can be reduced to one master equation
\begin{equation}
\label{master_equation}
\pounds_{K} \Theta = 2\Theta \Big( 3 \chi - \frac{\partial \delta^{\dot{N}}}{\partial q^{\dot{N}}} \Big)  + \mathcal{P}
\end{equation}
where $\mathcal{P}$ is the third-order polynomial in $p_{\dot{A}}$
\begin{equation}
\label{definicja_wielomianu_P}
\mathcal{P} := \frac{1}{6} \frac{\partial^{2} \delta_{\dot{A}}}{\partial q^{\dot{B}} \partial q^{\dot{C}}} \, p^{\dot{A}}p^{\dot{B}}p^{\dot{C}} + \Big( \frac{1}{3} F_{\dot{A}}\epsilon_{\dot{B}} + \frac{1}{2} \frac{\partial \epsilon_{\dot{A}}}{\partial q^{\dot{B}}} \Big) p^{\dot{A}}p^{\dot{B}} + \zeta_{\dot{A}} p^{\dot{A}} + \xi
\end{equation}

Integrability conditions are
\begin{subequations}
\begin{eqnarray}
\label{integrability_condition_1}
&&\Lambda \chi=0  \ \ \ \ \ , \ \ \ \ \chi=\chi(q^{\dot{M}})
\\ 
\label{integrability_condition_2}
&& C^{(2)} \, \frac{\partial \chi}{\partial q_{\dot{A}}} = 0 
\\
\label{integrability_condition_3}
&&-\Lambda \epsilon^{\dot{A}} = \delta^{\dot{A}} \frac{\partial F^{\dot{M}}}{\partial q^{\dot{M}}} + \frac{\partial}{\partial q_{\dot{A}}} \Big( \delta_{\dot{N}} F^{\dot{N}} + \frac{\partial \delta^{\dot{N}}}{\partial q^{\dot{N}}} - 3 \chi \Big)
\\
&&  \frac{\partial}{\partial q_{\dot{A}}} \big( \delta_{\dot{N}} Z^{\dot{N}} \big) + Z^{\dot{A}} \frac{\partial \delta^{\dot{N}}}{\partial q^{\dot{N}}} + \delta^{\dot{A}} \frac{\partial Z^{\dot{N}}}{\partial q^{\dot{N}}}=0  
\\
\label{integrability_condition_5}
&& \frac{\partial G^{\dot{A}}}{\partial q^{\dot{A}}} + F_{\dot{A}}G^{\dot{A}} - 2 N^{\dot{A}} \frac{\partial \chi}{\partial q^{\dot{A}}}=0
\\
\label{integrability_condition_6}
&&\frac{\partial \chi}{\partial q^{\dot{N}}} \frac{\partial \Theta}{\partial p_{\dot{N}}} = \Big( \frac{1}{2} \frac{\partial^{2} \chi}{\partial q^{\dot{A}} \partial q^{\dot{B}}} + \frac{1}{3} F_{\dot{A}} \frac{\partial \chi}{\partial q^{\dot{B}}} \Big) p^{\dot{A}} p^{\dot{B}} + \frac{1}{2} G_{\dot{A}} \, p^{\dot{A}} + G
\end{eqnarray}
\end{subequations}

where 
\begin{eqnarray}
\label{definicja_GA}
G^{\dot{A}} &:=& \frac{\partial}{\partial q_{\dot{A}}} \Big( 2\delta_{\dot{N}} N^{\dot{N}} - \frac{\partial \epsilon^{\dot{N}}}{\partial q^{\dot{N}}} \Big) + F^{\dot{A}} \, \frac{\partial \epsilon^{\dot{N}}}{\partial q^{\dot{N}}} + \epsilon^{\dot{A}} \, \frac{\partial F^{\dot{N}}}{\partial q^{\dot{N}}} 
\\ \nonumber
&&+ 2 \delta^{\dot{A}} \, \frac{\partial N^{\dot{N}}}{\partial q^{\dot{N}}} + 2 N^{\dot{A}} \, \frac{\partial \delta^{\dot{N}}}{\partial q^{\dot{N}}} - 4\chi N^{\dot{A}}
\\ \nonumber
&=&- \frac{\partial^{2} \epsilon^{\dot{M}}}{\partial q_{\dot{A}} \partial q^{\dot{M}}} + F^{\dot{A}} \frac{\partial \epsilon^{\dot{M}}}{\partial q^{\dot{M}}} + \epsilon^{\dot{A}} \frac{\partial F^{\dot{M}}}{\partial q^{\dot{M}}} + 2\, \frac{\partial}{\partial q^{\dot{M}}} \Big( \delta^{\dot{M}} N^{\dot{A}} \Big) + 2 N^{\dot{N}} \frac{\partial \delta_{\dot{N}}}{\partial q_{\dot{A}}} - 4 \chi N^{\dot{A}}
\\ 
\label{definicja_G}
G &:=& \frac{\partial \zeta^{\dot{A}}}{\partial q^{\dot{A}}} +F_{\dot{A}} \zeta^{\dot{A}} +N_{\dot{A}} \epsilon^{\dot{A}} +\Lambda \xi + \delta^{\dot{A}} \frac{\partial \gamma}{\partial q^{\dot{A}}} + 2 \gamma \, \frac{\partial \delta^{\dot{N}}}{\partial q^{\dot{N}}} - 4 \chi \gamma
\end{eqnarray}
and $Z^{\dot{A}}$ is given by (\ref{definicja_Z}).

\subsection{Transformation formulas.}

Under the gauge given by (\ref{gaug_wspolrzednych}), (\ref{transformacja_F}) - (\ref{transformacja_fun_klucz_dla_nonexpan}) and (\ref{definicje_funkcyj_w_gaugu_dla_nonexpan}), $\delta^{\dot{M}}$, $\epsilon^{\dot{M}}$, $\zeta^{\dot{M}}$ and $\xi$ transform according to
\begin{subequations}
\begin{eqnarray} 
\label{transformacja_delty_nieniebo}
\delta'^{\dot{M}} &=& \Delta \, D^{-1 \ \ \dot{M}}_{\ \ \; \dot{B}} \delta^{\dot{B}}
\\
\label{transformacja_epsilon_nieniebo}
\epsilon'^{\dot{M}} &=& D^{-1 \ \ \dot{M}}_{\ \ \; \dot{B}} \epsilon^{\dot{B}} - \sigma_{\dot{B}} \, \frac{\partial \delta'^{\dot{B}}}{\partial q'_{\dot{M}}} + \delta^{\dot{B}} \, \frac{\partial \sigma^{\dot{M}}}{\partial q^{\dot{B}}} -2\chi \, \sigma^{\dot{M}}
\\ \nonumber
 &=& D^{-1 \ \ \dot{M}}_{\ \ \; \dot{B}} \epsilon^{\dot{B}} - \frac{\partial}{\partial q'_{\dot{M}}} (\sigma_{\dot{B}} \delta'^{\dot{B}} ) + 2h \, \delta'^{\dot{M}} - 2 \chi \, \sigma^{\dot{M}}
\\
\label{transformacja_zeta_nieniebo}
\Delta^{2} \zeta'_{\dot{R}} &=&  D_{\dot{R}}^{\ \ \dot{N}} \bigg[ \zeta_{\dot{N}} - \delta^{\dot{M}} \frac{\partial H_{\dot{N}}}{\partial q^{\dot{M}}} + H_{\dot{M}} \frac{\partial \delta_{\dot{N}}}{\partial q_{\dot{M}}} +\epsilon^{\dot{M}} N_{\dot{M}\dot{N}} +2 H_{\dot{N}} \bigg( 2 \chi - \frac{\partial \delta^{\dot{M}}}{\partial q^{\dot{M}}} \bigg) \bigg] \ \ \ \ \ \ 
\\ \nonumber
&& -\frac{\Delta^{2}}{2} \frac{\partial^{2} \delta'_{(\dot{R}}}{\partial q'^{\dot{S}} \partial q'^{\dot{T})}} \, \sigma^{\dot{S}} \sigma^{\dot{T}} - \Delta^{2}\sigma^{\dot{M}} \bigg(  \frac{\partial \epsilon'_{(\dot{M}}}{\partial q'^{\dot{R})}} + \frac{2}{3} \,  F'_{(\dot{M}} \epsilon'_{\dot{R})} 
\bigg)
\\ 
\label{transformacja_xi_nieniebo}
\Delta^{2} \xi' &=& \xi + 2M \bigg( \frac{\partial \delta^{\dot{N}}}{\partial q^{\dot{N}}} - 3\chi \bigg) +\delta^{\dot{N}} \frac{\partial M}{\partial q^{\dot{N}}} - \epsilon^{\dot{N}} H_{\dot{N}}
\\ \nonumber
&& -\Delta^{2} \bigg[ \frac{1}{6} \, \frac{\partial^{2} \delta'_{\dot{R}}}{\partial q'^{\dot{S}} \partial q'^{\dot{T}}} \, \sigma^{\dot{R}} \sigma^{\dot{S}} \sigma^{\dot{T}} + \bigg( \frac{1}{2} \, \frac{\partial \epsilon'_{\dot{R}}}{\partial q'^{\dot{S}}} + \frac{1}{3} F'_{\dot{R}} \epsilon'_{\dot{S}} \bigg) \, \sigma^{\dot{R}} \sigma^{\dot{S}} + \zeta'_{\dot{R}} \, \sigma^{\dot{R}} \bigg]
\end{eqnarray}
\end{subequations}
In classification of the Killing vector, transformation formulas for spinorial divergences $\frac{\partial \delta^{\dot{M}}}{\partial q^{\dot{M}}}$ and $\frac{\partial \epsilon^{\dot{M}}}{\partial q^{\dot{M}}}$ will be useful
\begin{subequations}
\begin{eqnarray}
\label{divergencja_delty}
\frac{\partial \delta'^{\dot{M}}}{\partial q'^{\dot{M}}} &=& \frac{\partial \delta^{\dot{M}}}{\partial q^{\dot{M}}} + \delta^{\dot{M}} \, \frac{\partial \ln \Delta}{\partial q^{\dot{M}}}
\\ 
\label{divergencja_epsilona}
\frac{\partial \epsilon'^{\dot{M}}}{\partial q'^{\dot{M}}} &=& \Delta^{-1} \frac{\partial \epsilon^{\dot{M}}}{\partial q^{\dot{M}}} + 2 \delta^{\dot{M}} \, \frac{\partial h}{\partial q^{\dot{M}}} + 2h \bigg( \frac{\partial \delta'^{\dot{M}}}{\partial q'^{\dot{M}}} - 2 \chi \bigg)  - 2 \sigma^{\dot{M}} \frac{\partial \chi}{\partial q'^{\dot{M}}}
\end{eqnarray}
\end{subequations}

\renewcommand{\arraystretch}{1.5}
\setlength\arraycolsep{2pt}
\setcounter{equation}{0}

\section{Classification of the isometric and homothetic Killing vectors in nonexpanding hyperheavenly spaces.}
\label{sekcja_Killing}

\subsection{Criterion used in classification.}

Classification of the Killing symmetries that we propose, is based on the algebraic properties of the components of the Killing vector, $\delta^{\dot{M}}$ and $\epsilon^{\dot{M}}$. If the $\delta^{\dot{M}} \ne 0$, then from (\ref{divergencja_delty}) it follows, that $\frac{\partial \delta^{\dot{M}}}{\partial q^{\dot{M}}}$ can be always gauged away, bringing $\delta^{\dot{M}}$ to gradient. The appropriate gauge transformation allows then to bring $\delta^{\dot{M}}$ to the form $\delta^{\dot{M}}=\delta^{\dot{M}}_{\dot{Z}}$ with $\dot{Z}$ being some fixed index, $\dot{1}$ or $\dot{2}$. We always choose $\dot{Z}=\dot{1}$, namely $\delta^{\dot{M}}=\delta^{\dot{M}}_{\dot{1}}$. The second choice is completely analogous and does not lead to different results (since structure of the nonexpanding hyperheavenly space is invariant under the interchanging coordinates $q^{\dot{1}}$ and $q^{\dot{2}}$). This case we call type HK1 or IK1 (homothetic Killing vector of the type 1 or isometric Killing vector of the type 1, respectively). After setting $\delta^{\dot{A}} = \delta^{\dot{A}}_{\dot{1}}$ we still have the coordinate gauge transformations: $q'^{\dot{1}}= q^{\dot{1}} + f(q^{\dot{2}})$, $q'^{\dot{2}} = q'^{\dot{2}} (q^{\dot{2}})$ with $\Delta=\Delta(q^{\dot{2}}) = \frac{\partial q'^{\dot{2}}}{\partial q^{\dot{2}}}$ and $f$ being the arbitrary function.

Isometric symmetries with $\delta^{\dot{M}}=0$ have two subcases. Isometric type IK2a is characterized by $\frac{\partial \epsilon^{\dot{M}}}{\partial q^{\dot{M}}} \ne 0$. Using $\Delta$, this divergence can be bring to the constant value, say $\frac{\partial \epsilon^{\dot{M}}}{\partial q^{\dot{M}}} = 1$, (compare (\ref{divergencja_epsilona})). Since in two dimensions each vector is proportional to gradient, one can finally bring $\epsilon^{\dot{M}}$ to the form $\epsilon^{\dot{1}}= q^{\dot{1}}$, $\epsilon^{\dot{2}}=0$. In that case admissible coordinate gauge transformations are: $q'^{\dot{1}}=q^{\dot{1}} \, f(q^{\dot{2}})$, $q'^{\dot{2}} = q'^{\dot{2}} (q^{\dot{2}})$ with $\Delta=1$ and $\frac{\partial q'^{\dot{2}}}{\partial q^{\dot{2}}} = f^{-1}$. But if from the very beginning $\frac{\partial \epsilon^{\dot{M}}}{\partial q^{\dot{M}}} = 0$, without any loss of generality one can put $\epsilon^{\dot{M}} = \delta^{\dot{M}}_{\dot{1}}$. This case we call type IK2b. After that choice, we are left with $q'^{\dot{1}} = q'^{\dot{1}} (q^{\dot{M}})$, $q'^{\dot{2}} = q^{\dot{2}} + \textrm{const}$, $\Delta=\Delta(q^{\dot{M}}) = \frac{\partial q'^{\dot{1}}}{\partial q^{\dot{1}}}$. 

Homothetic symmetries with $\delta^{\dot{M}}=0$ do not involve two different types. Nonzero $\chi_{0}$ allows always gauge away $\epsilon^{\dot{M}}$. That is why this case we simply call type HK2. 

Gathering, we obtain the following types of the isometric and homothetic Killing vectors in nonexpanding hyperheavenly spaces
\newline
\newline
\begin{tabular}{|c|c|c|c|c|}   \hline
\multicolumn{3}{|c|}{Isometric Killing vectors} & \multicolumn{2}{|c|}{Homothetic Killing vectors}  \\  \hline
IK1 & IK2a & IK2b & HK1 & HK2 \\ \hline
$\delta^{\dot{A}}=\delta^{\dot{A}}_{\dot{1}}$ & $\delta^{\dot{A}}=0$ & $\delta^{\dot{A}}=0$ & $\delta^{\dot{A}}=\delta^{\dot{A}}_{\dot{1}}$ & $\delta^{\dot{A}}=0$  \\
$\epsilon^{\dot{A}}=0$  &  $\epsilon^{\dot{1}}= q^{\dot{1}}$, $\epsilon^{\dot{2}}=0$ & $\epsilon^{\dot{A}}=\delta^{\dot{A}}_{\dot{1}}$ & $\epsilon^{\dot{A}}=0$ & $\epsilon^{\dot{A}}=0$
\\ \hline
\end{tabular}
\newline
\newline

We do not use this way of classification in the case of the conformal symmetries. It seems that the way chosen in \cite{biblio_51} is the best description of the conformal symmetries. Moreover, the only non-conformally flat space which admits conformal symmetries is a space of the type $[\textrm{N}] \otimes [\textrm{N}]$. There is no need to classify one algebraic type.

The main aim of our considerations is to use coordinate gauge freedom and bring the master equation to the simplest possible form and then to solve it in order to obtain the form of the key function. The remain gauge freedom was used to simplify maximally the arbitrary structural functions $F^{\dot{A}}$, $N^{\dot{A}}$ and $\gamma$ appearing in the hyperheavenly equation. It is the main reason, why the form of the structural functions in the algebraic types $[\textrm{III,N}] \otimes [\textrm{any}]$ are sometimes different of the standard form obtained in many other works devoted to the problem of the nonexpading hyperheavenly spaces \cite{biblio_20, biblio_22, biblio_27}.

\subsection{Conformal Killing symmetries.}

Conformal Killing symmetries are allowed only when $\Lambda=0=C^{(2)}$, and $C_{\dot{A}\dot{B}\dot{C}\dot{D}} \frac{\partial \chi}{\partial q^{\dot{D}}}=0$, i.e. in the types $[\textrm{N},-] \otimes [\textrm{N},-]$. The $[\textrm{N}] \otimes [\textrm{N}]$ case was completely solved in \cite{biblio_51}.

\subsection{Homothetic Killing symmetries.}

We have here $\chi=\chi_{0}=\textrm{const}$ and we assume, that $\chi_{0} \ne 0$. From (\ref{integrability_condition_1}) we get immediately $\Lambda =0$. From (\ref{integrability_condition_6}) it follows that $G^{\dot{A}}=0=G$. The integrability conditions (\ref{integrability_condition_2}) and (\ref{integrability_condition_5}) are automatically satisfied. Thus one obtains the following cases:

\subsubsection{Type HK1 ($K=\partial_{q^{\dot{1}}} + 2\chi_{0} \, p^{\dot{B}} \partial_{p^{\dot{B}}} $)}

The forms of Killing vector and the key function are given by
\newline
\newline
\begin{tabular}{|c|c|c|c|}   \hline
Functions & Killing vector  & Master equation & The key function  \\  \hline
$\delta^{\dot{A}}=\delta^{\dot{A}}_{\dot{1}}$,  $\epsilon^{\dot{A}}=\zeta^{\dot{A}}=\xi=0$ & $K=\partial_{q^{\dot{1}}} + 2\chi_{0} \, p^{\dot{B}} \partial_{p^{\dot{B}}}$ & $\pounds_{K} \Theta=6\chi_{0} \Theta$   & $\Theta=e^{6 \chi_{0} q^{\dot{1}} } \, T(x,y,q^{\dot{2}})$ 
\\ \hline
\end{tabular}
\newline
\newline
where $x:=p^{\dot{1}} \, e^{-2 \chi_{0} q^{\dot{1}}}$ and $y:=p^{\dot{2}} \, e^{-2 \chi_{0} q^{\dot{1}}}$ and $T$ is an arbitrary function of its variables. With this Killing vector the integrability conditions of the master equation immediately give
\begin{equation}
F^{\dot{A}} = F^{\dot{A}} (q^{\dot{2}}) \ \ \ , \ \ \ N^{\dot{A}} = n^{\dot{A}} (q^{\dot{2}}) \, e^{2 \chi_{0} q^{\dot{1}}} \ \ \ , \ \ \ \gamma = g(q^{\dot{2}}) \, e^{4 \chi_{0} q^{\dot{1}}}
\end{equation}
where $n^{\dot{A}}$ and $g$ are arbitrary functions of their variable. The hyperheavenly equation takes the form
\begin{eqnarray}
&&T_{xx}T_{yy} - T_{xy}^{2} + 2\chi_{0} \big( 2 T_{y} - x \, T_{xy} - y \, T_{yy} \big) - T_{x q^{\dot{2}}} 
\\ \nonumber
&& + F^{\dot{1}} \bigg( T_{x} - \frac{2}{3} x \, T_{xx} - \frac{2}{3} y \, T_{xy} \bigg)
 + F^{\dot{2}} \bigg( T_{y} - \frac{2}{3} x \, T_{xy} - \frac{2}{3} y \, T_{yy} \bigg) 
\\ \nonumber
&&+ \frac{1}{18} \, \big( F^{\dot{1}}y - F^{\dot{2}} x \big)^{2} + \frac{1}{6} \bigg( \frac{\partial F^{\dot{2}}}{\partial q^{\dot{2}}} \, xy - \frac{\partial F^{\dot{1}}}{\partial q^{\dot{2}}} \, y^{2} \bigg) = n^{\dot{2}} x - n^{\dot{1}}y +g
\end{eqnarray}
The remaining gauge freedom can be employed in particular algebraic types:
\newline
\newline
\begin{tabular}{|c|c|}   \hline
 $[\textrm{III}] \otimes [\textrm{any}]$  &  $[\textrm{N}] \otimes [\textrm{any}]$   \\ \hline
$F^{\dot{1}}=0$, $F^{\dot{2}}=F_{0} q^{\dot{2}}$, $F_{0}=\textrm{const} \ne 0$, $F_{0}'=\Delta_{0}^{-1} F_{0}$
& $F^{\dot{1}}=0$, $F^{\dot{2}}=F_{0} =\textrm{const}$  \\
$n^{\dot{A}}=0$ &  $n^{\dot{1}}=0$, $n^{\dot{2}}=n(q^{\dot{2}})$  \\ 
$g=0$ & $g=0$  \\  \hline
$C^{(2)}=-F_{0}$ & $C^{(2)}=0$  \\
$C^{(1)} = -F_{0}^{2} \, q^{\dot{2}} p^{\dot{1}}$ & $C^{(1)} = -2 \, \frac{\partial N^{\dot{2}}}{\partial q^{\dot{2}}}$ \\ \hline
$l_{12} = -F_{0} \, q^{\dot{2}} - 2 \chi_{0}$ & $l_{12} = -F_{0} - 2\chi_{0}$  \\
$l_{22} = -F_{0} \, p^{\dot{2}}$ & $l_{22} =-2 N^{\dot{2}}$  \\ \hline
\end{tabular}
\newline
\newline
(where $\Delta_{0}$ is a constant determinant (\ref{definicja_wyznacznika_macierzy_przejscia}) and $n=n(q^{\dot{2}})$ is an arbitrary function of $q^{\dot{2}}$).

\subsubsection{Type HK2 ($K= 2\chi_{0} \, p^{\dot{B}} \partial_{p^{\dot{B}}} $)}

The forms of Killing vector and the key function are given by:
\newline
\newline
\begin{tabular}{|c|c|c|c|}   \hline
Functions & Killing vector  & Master equation & The key function  \\  \hline
$\delta^{\dot{A}}=\epsilon^{\dot{A}}=\zeta^{\dot{A}}=\xi=0$ & $K= 2\chi_{0} \, p^{\dot{B}} \partial_{p^{\dot{B}}}$ & $\pounds_{K} \Theta=6\chi_{0} \Theta$   & $\Theta= (p^{\dot{2}})^{3}  \, T ( x,q^{\dot{M}} ) $ 
\\ \hline
\end{tabular}
\newline
\newline 
$T$ is an arbitrary function of its variables and $x:= \frac{p^{\dot{1}}}{p^{\dot{2}}}$. 
\newline
Integrability conditions of the master equation give $N^{\dot{A}}=\gamma=0$. Moreover, $C^{(2)}$ must be nonzero; in the other case we obtain automatically a heavenly space. The only allowed algebraic type is $[\textrm{III}] \otimes [\textrm{any}]$ in which the structural functions can be brought to the form
\newline
\newline
\begin{tabular}{|c|}   \hline
 $[\textrm{III}] \otimes [\textrm{any}]$     \\ \hline
$F^{\dot{1}}=F_{0} q^{\dot{1}}$, $F^{\dot{2}}=0$, $F_{0}=\textrm{const} \ne 0$, $F_{0}'=\Delta_{0}^{-1} F_{0}$ \\
$N^{\dot{A}}=0$  , $\gamma=0$  \\  \hline
$C^{(2)}=-F_{0}$   , $C^{(1)} = F_{0}^{2} \, q^{\dot{1}} p^{\dot{2}}$  \\ \hline
$l_{12} = -2\chi_{0}$  , $l_{22} =0$  \\ \hline
\end{tabular}
\newline
\newline
The hyperheavenly equation become then
\begin{equation}
\label{rownanie_hiperniebianskie_w_typie_HK2}
6 T T_{xx} -4  T_{x}^{2} - T_{x q^{\dot{2}}} - 3 T_{q^{\dot{1}}} -x \, T_{x q^{\dot{1}}}
 -\frac{1}{3} F_{0} q^{\dot{1}} \,  T_{x}  +\frac{1}{18} \big( F_{0} \, q^{\dot{1}} \big)^{2} - \frac{1}{6} F_{0} \, x = 0
\end{equation}

\subsection{Isometric Killing symmetries with $\Lambda \ne 0$.}

Here we have $\chi=0$. From (\ref{integrability_condition_6}) it follows, that $G^{\dot{A}}=0=G$. Equations  (\ref{integrability_condition_2}) and (\ref{integrability_condition_5}) reduce to the identities $0=0$. Nonzero $\Lambda$ means, that $C^{(3)} \ne 0$ and the only allowed types are $[\textrm{II,D}] \otimes [\textrm{any}]$. In both types $F^{\dot{A}}$ can be gauged away what we assume to be done.

\subsubsection{Type IK1 ($K=\partial_{q^{\dot{1}}}$)}

The forms of Killing vector and the key function are presented by the table:
\newline
\newline
\begin{tabular}{|c|c|c|c|}   \hline
Functions & Killing vector  & Master equation & The key function  \\  \hline
$\delta^{\dot{A}}=\delta^{\dot{A}}_{\dot{1}}$,  $\epsilon^{\dot{A}}=\zeta^{\dot{A}}=\xi=0$ & $K=\partial_{q^{\dot{1}}}$ & $\pounds_{K} \Theta=0$   & $\Theta=\Theta (q^{\dot{2}}, p^{\dot{M}})$ 
\\ \hline
\end{tabular}
\newline
\newline
With $F^{\dot{A}}=0$ the integrability conditions of the master equation give $N^{\dot{A}}=N^{\dot{A}} (q^{\dot{2}})$ and $\gamma=\gamma(q^{\dot{2}})$. The hyperheavenly equation takes the form
\begin{equation}
\frac{1}{2} \, \Theta_{p_{\dot{A}} p_{\dot{B}}} \Theta_{p^{\dot{A}} p^{\dot{B}}}
- \Theta_{p^{\dot{1}} q^{\dot{2}}} 
+\Lambda \Big( p^{\dot{A}} \Theta_{p^{\dot{A}}} - \Theta - \frac{1}{3} \, p^{\dot{A}} p^{\dot{B}} \Theta_{p^{\dot{A}} p^{\dot{B}}}   \Big)
= N_{\dot{A}} \, p^{\dot{A}} +\gamma \ \ \ \ \ \ \ \ \ 
\end{equation}
The type $[\textrm{D}] \otimes [\textrm{any}]$ condition, $2 C^{(2)}C^{(2)}-3 C^{(1)}C^{(3)}=0$ is equivalent to $Z^{\dot{A}}=0$. It means, that for the type $[\textrm{D}] \otimes [\textrm{any}]$, $F^{\dot{A}}=0$ implies $N^{\dot{A}}=0$. For the type $[\textrm{II}] \otimes [\textrm{any}]$, $N^{\dot{A}}$ must be nonzero but there appear two different ways for further simplification:
\newline
\newline
\begin{tabular}{|c|c|c|}   \hline
 $[\textrm{D}] \otimes [\textrm{any}]$  &  \multicolumn{2}{|c|}{$[\textrm{II}] \otimes [\textrm{any}]$}   \\ \hline
$F^{\dot{A}}=0$ & $F^{\dot{A}}=0$ & $F^{\dot{A}}=0$ \\
$N^{\dot{A}}=0$ &  $N^{\dot{1}}=0$, $N^{\dot{2}}=N_{0} = \textrm{const} \ne 0$   
& $N^{\dot{2}}=0$, $N^{\dot{1}}=N_{0} = \textrm{const}$  \\ 
  & $N_{0}'=\Delta_{0}^{-1} N_{0}$ & $N_{0}'=\Delta_{0}^{-2} N_{0}$  \\
$\gamma=0$ & $\gamma=0$ & $\gamma=0$ \\  \hline
$C^{(3)}=-\frac{2}{3} \Lambda$ & $C^{(3)}=-\frac{2}{3} \Lambda$ & $C^{(3)}=-\frac{2}{3} \Lambda$  \\
$C^{(2)}=0$ & $C^{(2)}=0$ & $C^{(2)}=0$  \\
$C^{(1)} = 0$ & $C^{(1)} = 2 \Lambda N_{0} \, p^{\dot{1}}$ & $C^{(1)} = -2 \Lambda N_{0} \, p^{\dot{2}}$ \\ \hline
$l_{12} = -\Lambda \, p^{\dot{2}} $ & $l_{12} = -\Lambda \, p^{\dot{2}} $ & $l_{12} = -\Lambda \, p^{\dot{2}} $  \\
$l_{22} = 0$ & $l_{22} =-2 N_{0}$  & $l_{22}=0$ \\ \hline
\end{tabular}

\subsubsection{Type IK2 }

This type is not allowed in this case, since from (\ref{integrability_condition_3}) with $\delta^{\dot{A}}=0$ it immediately follows that $\epsilon^{\dot{A}}=0$. Therefore, when $\Lambda \ne 0$, there is no Killing vector of the type IK2.

\subsection{Isometric Killing symmetries with $\Lambda = 0$.}

As in the previous case, $\chi=0$, $G^{\dot{A}}=0=G$, (\ref{integrability_condition_2}) and (\ref{integrability_condition_5}) reduce to the identities $0=0$. However, $\Lambda=0$ implies the types $[\textrm{III,N}] \otimes [\textrm{any}]$.

\subsubsection{Type IK1 ($K=\partial_{q^{\dot{1}}}$)}

The forms of Killing vector and the key function are given by:
\newline
\newline
\begin{tabular}{|c|c|c|c|}   \hline
Functions & Killing vector  & Master equation & The key function  \\  \hline
$\delta^{\dot{A}}=\delta^{\dot{A}}_{\dot{1}}$,  $\epsilon^{\dot{A}}=\zeta^{\dot{A}}=\xi=0$ & $K=\partial_{q^{\dot{1}}} $ & $\pounds_{K} \Theta=0$   & $\Theta=\Theta(q^{\dot{2}},p^{\dot{M}})$ 
\\ \hline
\end{tabular}
\newline
\newline
The integrability conditions of the master equation gives $F^{\dot{A}} = F^{\dot{A}} (q^{\dot{2}})$, $N^{\dot{A}} = N^{\dot{A}} (q^{\dot{2}})$, $\gamma=\gamma(q^{\dot{2}})$. Concrete algebraic types are characterized by:
\newline
\newline
\begin{tabular}{|c|c|c|}   \hline
 $[\textrm{III}] \otimes [\textrm{any}]$  &  \multicolumn{2}{|c|}{$[\textrm{N}] \otimes [\textrm{any}]$}   \\ \hline
$F^{\dot{1}}=0$, $F^{\dot{2}}=F_{0} q^{\dot{2}}$, & $F^{\dot{1}}=0$, $F^{\dot{2}}=F_{0} =\textrm{const} \ne 0$ & $F^{\dot{A}}=0$ \\
$F_{0}=\textrm{const} \ne 0$, $F_{0}'=\Delta_{0}^{-1} F_{0}$ & & \\
$N^{\dot{A}}=0$ &  $N^{\dot{2}}=0$, $N^{\dot{1}}=N_{0}=\textrm{const} \ne 0$ & $N^{\dot{1}}=0$, 
$N^{\dot{2}}=N^{\dot{2}}(q^{\dot{2}}) $  \\ 
$\gamma=0$ & $\gamma=0$ & $\gamma=0$  \\  \hline
$C^{(2)}=-F_{0}$ & $C^{(2)}=0$ & $C^{(2)}=0$ \\
$C^{(1)} = -F_{0}^{2} q^{\dot{2}} p^{\dot{1}}$ & $C^{(1)}=-2 F_{0} N_{0}$ & $C^{(1)} = -2 \, \frac{\partial N^{\dot{2}}}{\partial q^{\dot{2}}}$ \\ \hline
$l_{12} = -F_{0} \, q^{\dot{2}} $ & $l_{12} = -F_{0} $ & $l_{12}=0$  \\
$l_{22} = -F_{0} \, p^{\dot{2}}$ & $l_{22}=0$ & $l_{22} =-2 N^{\dot{2}}$  \\ \hline
\end{tabular}

\subsubsection{Type IK2a ($K= q^{\dot{1}} \partial_{p^{\dot{1}}}$)}

The forms of Killing vector and the key function read:
\newline
\newline
\begin{tabular}{|c|c|c|c|}   \hline
Functions & Killing vector  & Master equation & The key function  \\  \hline
$\delta^{\dot{A}}=\zeta^{\dot{A}}=\xi=0$ & $K= q^{\dot{1}} \partial_{p^{\dot{1}}} $ & $\pounds_{K} \Theta=$   & $\Theta=T(q^{\dot{M}},p^{\dot{2}})$ \\
$\epsilon^{\dot{1}}= q^{\dot{1}}$, $\epsilon^{\dot{2}}=0$ &  & $\frac{1}{3} F_{0} \, p^{\dot{2}}p^{\dot{2}} - \frac{1}{2} \,  p^{\dot{1}}p^{\dot{2}} $ & $+\frac{1}{q^{\dot{1}}} \Big( \frac{1}{3} F_{0} \, p^{\dot{1}}p^{\dot{2}}p^{\dot{2}} - \frac{1}{4} \, p^{\dot{1}}p^{\dot{1}}p^{\dot{2}} \Big)$
\\ \hline 
\end{tabular}
\newline
\newline
The integrability conditions of the master equation allow to find the forms of the $F^{\dot{A}}$ and $N^{\dot{A}}$. Concrete algebraic types are characterized by:
\newline
\newline
\begin{tabular}{|c|c|}   \hline
 $[\textrm{III}] \otimes [\textrm{any}]$  &  $[\textrm{N}] \otimes [\textrm{any}]$   \\ \hline
$F^{\dot{1}}=F_{0} \, \frac{1}{q^{\dot{1}}}$, $F^{\dot{2}}=0$, $F_{0}=\textrm{const} \ne 0$, $F_{0}'= f_{0} F_{0}$
& $F^{\dot{A}}=0 \rightarrow F_{0}=0$  \\
$N^{\dot{2}}=0$, $N^{\dot{1}}=N^{\dot{1}}(q^{\dot{M}})$ &  $N^{\dot{2}}=0$, $N^{\dot{1}}=N^{\dot{1}}(q^{\dot{M}}) $  \\ 
$\gamma=0$ & $\gamma=0$  \\  \hline
$C^{(2)}=F_{0} \, \frac{1}{(q^{\dot{1}})^{2}}$ & $C^{(2)}=0$  \\
$C^{(1)} = -2 \, \frac{\partial N^{\dot{1}}}{\partial q^{\dot{1}}} -2 F_{0} \, \frac{p^{\dot{1}}}{(q^{\dot{1}})^{3}} -F_{0}^{2} \, \frac{p^{\dot{2}}}{(q^{\dot{1}})^{3}} $ & $C^{(1)} = -2 \, \frac{\partial N^{\dot{1}}}{\partial q^{\dot{1}}}$ \\ \hline
$l_{12} = 0 $ & $l_{12} =0 $  \\
$l_{22} = -1 $ & $l_{22} =-1$  \\ \hline
\end{tabular}
\newline
\newline
$f_{0}$ is an arbitrary complex, gauge constant, which can be used in order to bring $F_{0}$ to $1$ if desired. 
The hyperheavenly equation takes the form
\begin{equation}
\label{zredukowane_rownanie_hiperniebianskie_typ_IK2a}
-\frac{1}{2} \frac{p^{\dot{2}}}{q^{\dot{1}}} \, T_{p^{\dot{2}}p^{\dot{2}}} + T_{p^{\dot{2}}q^{\dot{1}}} - \frac{1}{2} F_{0}^{2} \bigg(\frac{p^{\dot{2}}}{q^{\dot{1}}} \bigg)^{2} + N^{\dot{1}} p^{\dot{2}} = 0
\end{equation}

\subsubsection{Type IK2b ($K=  \partial_{p^{\dot{1}}}$)}

The forms of Killing vector and the key function are:
\newline
\newline
\begin{tabular}{|c|c|c|c|}   \hline
Functions & Killing vector  & Master equation & The key function  \\  \hline
$\epsilon^{\dot{A}}=\delta^{\dot{A}}_{\dot{1}}$, $\delta^{\dot{A}}=\zeta^{\dot{A}}=\xi=0$ & $K=  \partial_{p^{\dot{1}}}$ & $\pounds_{K} \Theta=0$   & $\Theta=  \Theta(p^{\dot{2}},q^{\dot{M}})$ 
\\ \hline
\end{tabular}
\newline
\newline
Integrability conditions of the master equation gives $C^{(2)}=0$, so the only allowed algebraic type is $[\textrm{N}] \otimes [\textrm{any}]$. Coordinate gauge freedom can be used to set:
\newline
\newline
\begin{tabular}{|c|}   \hline
 $[\textrm{N}] \otimes [\textrm{any}]$     \\ \hline
$F^{\dot{A}}=0$, $N^{\dot{2}}=0$, $N^{\dot{1}}=N^{\dot{1}} (q^{\dot{M}})$, $\gamma=0$  \\  \hline
$C^{(2)}=0$ , $C^{(1)} = -2 \, \frac{\partial N^{\dot{1}}}{\partial q^{\dot{1}}}$  \\ \hline
$l_{12} = 0$ , $l_{22} =0$  \\ \hline
\end{tabular}
\newline
\newline
The hyperheavenly equation takes a very simple form
\begin{equation}
\label{zredukowane_rownanie_hiperniebianskie_typ_IK2b}
\Theta_{p^{\dot{2}} q^{\dot{1}}} = -N^{\dot{1}} p^{\dot{2}}
\end{equation}

\subsection{Examples of the nonexpanding hyperheavenly spaces admitting Killing vector. Lorentzian real slices.}

Existence of the Killing vector of the form $K= q^{\dot{1}} \partial_{p^{\dot{1}}}$ (type IK2a) or 
$K=\partial_{p^{\dot{1}}}$ (type IK2b) assures, that nonlinearities in hyperheavenly equation (\ref{nonexpanding_hyperheavenly_equation}) disappear. In that cases hyperheavenly equation becomes linear differential equation and it can be easily solved.

\subsubsection{Type IK2a.}

The hyperheavenly equation reduces to the form (\ref{zredukowane_rownanie_hiperniebianskie_typ_IK2a}) (if $F_{0}=0$ the algebraic type is $[\textrm{N}] \otimes [\textrm{any}]$). Writing $N^{\dot{1}}$ in the form
\begin{equation}
N^{\dot{1}} = \frac{1}{2 q^{\dot{1}}} \, N - \frac{\partial N}{\partial q^{\dot{1}}}
\end{equation}
with $N$ being the arbitrary function of the $q^{\dot{M}}$, one can find the solution for $T$ and finally, for $\Theta$
\begin{equation}
\Theta = p^{\dot{2}} \, S(q^{\dot{2}}, q^{\dot{1}} p^{\dot{2}} p^{\dot{2}}) 
-\frac{1}{12} \frac{p^{\dot{2}}}{q^{\dot{1}}} (F_{0} \, p^{\dot{2}} - p^{\dot{1}})(F_{0} \, p^{\dot{2}} - 3p^{\dot{1}}) + \frac{1}{2} N \, (p^{\dot{2}})^{2} + r(q^{\dot{M}})
\end{equation}
where $S=S(q^{\dot{2}}, q^{\dot{1}} p^{\dot{2}} p^{\dot{2}})$ is an arbitrary function of its variables. Arbitrary function $r=r(q^{\dot{M}})$ can be gauged to 0 by using the gauge function $M$ - see the transformation formula (\ref{transformacja_fun_klucz_dla_nonexpan})). The only nonzero curvature coefficients read
\begin{eqnarray}
\label{krzywizna_dla_pierwszego_przykladu}
C^{(2)} &=& F_{0} \, \frac{1}{q^{\dot{1}}q^{\dot{1}}} 
\\ \nonumber
C^{(1)} &=& -2 \, \frac{\partial}{\partial q^{\dot{1}}} \bigg( \frac{1}{2 q^{\dot{1}}} \, N - \frac{\partial N}{\partial q^{\dot{1}}} \bigg) - 2 F_{0} \, \frac{p^{\dot{1}}}{(q^{\dot{1}})^{3}} - F_{0}^{2} \, \frac{p^{\dot{2}}}{(q^{\dot{1}})^{3}}
\\ \nonumber
\dot{C}^{(1)} &=& 2 \, \Big( p^{\dot{2}} \, S(q^{\dot{2}}, q^{\dot{1}} p^{\dot{2}} p^{\dot{2}}) \Big)_{p^{\dot{2}}p^{\dot{2}}p^{\dot{2}}p^{\dot{2}}}
\end{eqnarray}
From (\ref{krzywizna_dla_pierwszego_przykladu}) it follows, that the only allowed algebraic types are $[\textrm{III,N}] \otimes [\textrm{N}]$
\begin{eqnarray}
ds^{2} &=& -2 \,  dp^{\dot{A}} \underset{s}{\otimes} dq_{\dot{A}}  
           -2 \, \bigg( (p^{\dot{2}} \, S)_{p^{\dot{2}}p^{\dot{2}}}- \frac{F_{0}^{2}}{2} \, \frac{p^{\dot{2}}}{q^{\dot{1}}} + N  \bigg) \, dq_{\dot{1}} \underset{s}{\otimes} dq_{\dot{1}}
\\ \nonumber
           &&+ \frac{p^{\dot{2}}}{ q^{\dot{1}}} \, dq_{\dot{2}} \underset{s}{\otimes} dq_{\dot{2}}
           +2 \bigg( 2F_{0} \, \frac{p^{\dot{2}}}{q^{\dot{1}}} -  \frac{p^{\dot{1}}}{q^{\dot{1}}} \bigg) \, dq_{\dot{1}} \underset{s}{\otimes} dq_{\dot{2}}
\end{eqnarray}

\subsubsection{Type IK2b.}

Writing $N^{\dot{1}}$ in the form $N^{\dot{1}} = - \frac{\partial N}{\partial q^{\dot{1}}}$ with $N=N(q^{\dot{M}})$ we obtain the solution of Eq. (\ref{zredukowane_rownanie_hiperniebianskie_typ_IK2b})
\begin{equation}
\Theta = \frac{1}{2} N \, p^{\dot{2}} p^{\dot{2}} + A(p^{\dot{2}}, q^{\dot{2}}) + r(q^{\dot{M}})
\end{equation}
where $A=A(p^{\dot{2}}, q^{\dot{2}})$ is an arbitrary function of its variables. (As before, arbitrary function $r(q^{\dot{M}})$ can be gauged away). The metric has the form
\begin{equation}
\label{metryka_dla_zespolonej_pp_fali}
ds^{2} = -2 \,  dp^{\dot{A}} \underset{s}{\otimes} dq_{\dot{A}}  
          -2 \, (N+A_{p^{\dot{2}}p^{\dot{2}}}) \, dq_{\dot{1}} \underset{s}{\otimes} dq_{\dot{1}}
\end{equation}
Since
\begin{equation}
C^{(1)} = 2 \, \frac{\partial^{2} N}{\partial q^{\dot{1}}\partial q^{\dot{1}}} \ \ \ \ , \ \ \ \ 
\dot{C}^{(1)} = 2 A_{p^{\dot{2}}p^{\dot{2}}p^{\dot{2}}p^{\dot{2}}}
\end{equation}
the space considered can be of types $[\textrm{N},-] \otimes [\textrm{N},-]$.

The Lorentzian real slices of the complex spacetime can exist only if there exists a coordinate frame such that $\bar{C}_{ABCD}=C_{\dot{A}\dot{B}\dot{C}\dot{D}}$, where the overbar denotes complex conjugation (see \cite{biblio_17}). Here we have to assure that $\bar{C}^{(1)} = \dot{C}^{(1)}$ and surprisingly this condition can be easily investigated. Denoting
\begin{eqnarray}
\label{podstawienia}
&& p^{\dot{1}} \equiv v \ \ \ , \ \ \ 
p^{\dot{2}} \equiv \bar{\zeta} \ \ \ , \ \ \ 
q^{\dot{1}} \equiv \zeta \ \ \ , \ \ \ 
q^{\dot{2}} \equiv u 
\\ \nonumber
&&
N(q^{\dot{1}}, q^{\dot{2}}) \equiv f(\zeta,u) \ \ \ , \ \ \ A_{p^{\dot{2}}p^{\dot{2}}} (p^{\dot{2}}, q^{\dot{2}}) \equiv \bar{f}(\bar{\zeta},u)
\end{eqnarray}
we automatically obtain $\bar{C}^{(1)} = \dot{C}^{(1)}$. Of course, $\bar{C}_{ABCD}=C_{\dot{A}\dot{B}\dot{C}\dot{D}}$ are not the only conditions which give the real spacetime with Lorentzian signature. However, in this particular case the metric (\ref{metryka_dla_zespolonej_pp_fali}) becomes
\begin{equation}
\label{metryka_dla_pp_fali_lorentzowskiej}
ds^{2} = 2 \,  (d \zeta \underset{s}{\otimes} d \bar{\zeta} - du \underset{s}{\otimes} dv)  
          -2 \, H(\zeta, \bar{\zeta}, u) \, du \underset{s}{\otimes} du
\end{equation}
with $H(\zeta, \bar{\zeta}, u) = f(\zeta,u) + \bar{f}(\bar{\zeta},u)$, so (\ref{metryka_dla_pp_fali_lorentzowskiej}) is the metric for the pp-wave solution with Einstein field equation assumed \cite{biblio_52}. With (\ref{podstawienia}) assumed the key function takes the form
\begin{equation}
\label{funkcja_kluczowa_dla_pp_wave}
\Theta = \frac{1}{2} \, f(\zeta, u) \, \bar{\zeta}^{2} + \int \!\!\! \int \bar{f}(\bar{\zeta},u) \, d \bar{\zeta} d \bar{\zeta}
\end{equation}
In the hyperheavenly language, (\ref{funkcja_kluczowa_dla_pp_wave}) is the key function for the pp-wave solution.

It is worth-while to note, that the type IK2a does not have any Lorentzian real slice (compare with  \cite{biblio_52}).

\subsection{Null Killing Vector Fields.}

The Killing vector field is null when $0=K^{a}K_{a} = 2 k_{\dot{B}} h^{\dot{B}}$. Using (\ref{componentss_off_the_Killing_vector}) one gets
\begin{equation}
\label{zerowosc_wektora_Killinga}
0=\delta_{\dot{A}} \bigg( \delta_{\dot{S}} Q^{\dot{S}\dot{A}} - 2\chi_{0} \, p^{\dot{A}} - \frac{\partial \delta^{\dot{N}}}{\partial q_{\dot{A}}} \, p_{\dot{N}} - \epsilon^{\dot{A}} \bigg)
\end{equation}
There are two possibilities: $\delta_{\dot{A}}=0$ and $\delta_{\dot{A}} \ne 0$, additionally we assume, that $\chi=\chi_{0}=\textrm{const}$.

\subsubsection{$\delta_{\dot{A}} \ne 0$}

Here we deal with types IK1 and HK1. After using the gauge freedom to set $\delta^{\dot{A}}=\delta^{\dot{A}}_{\dot{1}}$, $\epsilon^{\dot{A}}=\zeta^{\dot{A}}=\xi=\gamma=0$, the condition (\ref{zerowosc_wektora_Killinga}) gives $0=Q^{\dot{2}\dot{2}}+2\chi_{0}p^{\dot{2}}$ and, finally, solution for the key function
\begin{equation}
\Theta = \frac{1}{2} \bigg( \frac{\Lambda}{3} p^{\dot{2}}p^{\dot{2}} + 2\chi_{0} p^{\dot{2}} + \frac{2}{3} F^{\dot{2}} p^{\dot{2}} \bigg) p^{\dot{1}}p^{\dot{1}} + \mathcal{S}(p^{\dot{2}}, q^{\dot{M}}) p^{\dot{1}} + \mathcal{U} (p^{\dot{2}}, q^{\dot{M}})
\end{equation}
with $\mathcal{S} = \mathcal{S}(p^{\dot{2}}, q^{\dot{M}})$ and $\mathcal{U} = \mathcal{U}(p^{\dot{2}}, q^{\dot{M}})$ being the arbitrary functions. Inserting this key function into master equation and its integrability conditions, and then into hyperheavenly equation we obtain the polynomials in $p^{\dot{1}}$. Using remaining gauge freedom to set $F^{\dot{1}}=0=N^{\dot{1}}$ we easily find, that solution gives the space of the type $[-] \otimes [\textrm{any}]$, i.e., the heavenly space. Summing up: there is no null Killing vector fields with nonzero $\delta_{\dot{A}}$ in nonexpanding hyperheavenly space of the type $[\textrm{II,D,III,N}] \otimes [\textrm{any}]$.

\subsubsection{$\delta_{\dot{A}} = 0$}

When $\delta^{\dot{A}}=0$, the condition (\ref{zerowosc_wektora_Killinga}) is automatically satisfied. Killing vector lies on null string and is null from definition. In that case we deal with the types HK2, IK2a and IK2b. IK2a and IK2b are solved completely in previous subsection. Hyperheavenly equation in the type HK2 reduces to the equation (\ref{rownanie_hiperniebianskie_w_typie_HK2}) and we are going to investigate this equation soon.

\renewcommand{\arraystretch}{1.7}
\setlength\arraycolsep{2pt}
\setcounter{equation}{0}

\section{Classification of the homothetic Killing vectors in heavenly spaces.}

\subsection{General considerations.}

Isometric Killing vectors in heavenly spaces were considered and classified in details in \cite{biblio_16}. Here we present the classification of the homothetic Killing vectors. 

The heavenly spaces can be easily obtained from the hyperheavenly spaces, by setting $C^{(1)}=C^{(2)}=C^{(3)}=0$. This gives
\begin{equation}
\Lambda= \frac{\partial F^{\dot{A}}}{\partial q^{\dot{A}}} = F^{\dot{A}}N_{\dot{A}}-\frac{\partial N^{\dot{A}}}{\partial q^{\dot{A}}}=0
\end{equation}
By using the gauge freedom in $\Delta$ and $h$ the functions $F^{\dot{A}}$ and $N^{\dot{A}}$ can be gauged away. In order not to generate the nonzero $F^{\dot{A}}$ and $N^{\dot{A}}$ we have to keep $\Delta=\Delta_{0}=\textrm{const}$ and $h=h_{0}=\textrm{const}$. Function $\gamma$ can be removed from considerations by using arbitrary gauge function $H^{\dot{A}}$ (compare (\ref{transformacja_gammy_prostrzej})). After that steps the hyperheavenly equation reduces to the well-known, second heavenly equation \cite{biblio_11}
\begin{equation}
\frac{1}{2} \, \Theta_{p_{\dot{A}} p_{\dot{B}}} \Theta_{p^{\dot{A}} p^{\dot{B}}}
+ \Theta_{p_{\dot{A}} q^{\dot{A}}} =0
\end{equation}
In \cite{biblio_51} the problem of Killing vectors in heavenly spaces was generalized for the case of the most general conformal factor in (\ref{podstawowe_nasze_rownania_Killinga}). It has been proved, that $\chi$ can be at most a linear function of the $p^{\dot{A}}$ coordinates and it has the form
\begin{equation}
\chi= \frac{\partial \alpha}{\partial q^{\dot{A}}} \, p^{\dot{A}} + \chi_{1}
\end{equation}
with $\alpha=\alpha(q^{\dot{M}})$ and $\chi_{1}=\chi_{1}(q^{\dot{M}})$ being arbitrary functions of their variables. Since we are interested in homothetic symmetries it is enough to set $\chi_{1}=\chi_{0}=\textrm{const}$ and $\alpha=\alpha_{0}=\textrm{const}$. Moreover, it has been proved in \cite{biblio_16} that $\alpha_{0}$ can be put zero without any loss of generality. [It is essential for our considerations, because nonzero $\alpha_{0}$ provides the factor with the first integral of the heavenly equation into the master equation. In that case analysis is much more difficult, see \cite{biblio_51, biblio_16}]. Detailed analysis show, that with the choice $\alpha_{0}=0$, $\chi=\chi_{0}=\textrm{const}$ all formulas presented in section \ref{sekcja_Killing} can be used. Especially important for further considerations are integrability conditions of the master equation
\begin{equation}
\frac{\partial \zeta^{\dot{A}}}{\partial q^{\dot{A}}} = 0 \ \ \ , \ \ \ \frac{\partial \delta^{\dot{A}}}{\partial q^{\dot{A}}} = \delta_{0} = \textrm{const} \ \ \ , \ \ \ 
\frac{\partial \epsilon^{\dot{A}}}{\partial q^{\dot{A}}} = - \epsilon_{0} = \textrm{const}
\end{equation}
with the transformation laws (compare (\ref{divergencja_delty}) and (\ref{divergencja_epsilona})):
\begin{eqnarray}
\delta'_{0} &=& \delta_{0}
\\ 
\epsilon'_{0} &=& \Delta_{0}^{-1} \epsilon_{0} - 2h_{0} \, (\delta_{0}-2 \chi_{0})
\end{eqnarray}

The most general type appears when both $\delta_{0}$ and $\epsilon_{0}$ are non-zero. Note, that it involves $\delta_{0}=2 \chi_{0}$ (in the other way $\epsilon_{0}$ could be always gauged to zero). In that case it is convenient to use $\Delta_{0}$ and set $\epsilon_{0}=-4 \chi_{0}$. This case we call $\mathcal{H}$HKI (\textsl{heavenly homothetic Killing vector of the type I}). 

Second type, $\mathcal{H}$HKII is characterized by $\delta_{0} \ne 0$ and $\epsilon_{0}=0$. Then $\delta_{0}$ can be different from $2\chi_{0}$ (in that case $\epsilon_{0}$, if nonzero, can be always gauged away) or equal $2 \chi_{0}$ (because there always exists a possibility, that $\epsilon_{0}$ is equal zero from the very beginning).

Third type, $\mathcal{H}$HKIII appears when $\delta_{0}=0=\epsilon_{0}$. There are two subcases: type $\mathcal{H}$HKIIIa, with $\delta^{\dot{A}} \ne 0$ and type $\mathcal{H}$HKIIIb for which $\delta^{\dot{A}} = 0$. 

Remaining gauge freedom allows us to put $\zeta^{\dot{A}} = \frac{\partial \zeta}{\partial q_{\dot{A}}}$ and $\xi=0$, and brings $\delta^{\dot{A}}$ and $\epsilon^{\dot{A}}$ to the most plausible form. Gathering, we obtain the following types of the homothetic Killing vectors in heavenly spaces
\newline
\newline
\begin{tabular}{|c|c|c|c|}   \hline
 $\mathcal{H}$HKI & $\mathcal{H}$HKII & $\mathcal{H}$HKIIIa &  $\mathcal{H}$HKIIIb \\  \hline
$\delta_{0} \ne 0$, $\delta_{0}=2\chi_{0}$ & $\delta_{0} \ne 0$ & $\delta_{0} = 0$ & $\delta_{0} = 0$  \\
$\epsilon_{0}\ne 0 $, $\epsilon_{0}=-4 \chi_{0}$  & $\epsilon_{0}=0$ & $\epsilon_{0}= 0$ & $\epsilon_{0}=0$ \\
 & & $\delta^{\dot{A}} \ne 0$ & $\delta^{\dot{A}} =0$ \\ \hline
$\delta^{\dot{1}}=2\chi_{0} \, q^{\dot{1}}$, $\delta^{\dot{2}}=0$ & $\delta^{\dot{1}}=\delta_{0} \, q^{\dot{1}}$, $\delta^{\dot{2}}=0$ & $\delta^{\dot{A}}=\delta^{\dot{A}}_{\dot{1}}$ & $\delta^{\dot{A}}=0$
\\ 
$\epsilon^{\dot{A}}=2\chi_{0} q^{\dot{A}}$ & $\epsilon^{\dot{A}}=0$ & $\epsilon^{\dot{A}}=0$ & $\epsilon^{\dot{A}}=0$
\\
$\zeta^{\dot{A}}=\xi=0$ & $\zeta^{\dot{A}}=\xi=0$ & $\zeta^{\dot{A}}=\xi=0$ & $\zeta^{\dot{A}}=\xi=0$ 
\\ \hline
$K=2\chi_{0} \Big( q^{\dot{1}} \frac{\partial}{\partial q^{\dot{1}}}   $ & 
$K=\delta_{0} q^{\dot{1}} \frac{\partial}{\partial q^{\dot{1}}} $ & 
$K=\frac{\partial}{\partial q^{\dot{1}}} + 2\chi_{0} p^{\dot{A}} \frac{\partial}{\partial p^{\dot{A}}}$ &
$K= 2\chi_{0} p^{\dot{A}} \frac{\partial}{\partial p^{\dot{A}}}$
\\
$+ p^{\dot{1}} \frac{\partial}{\partial p^{\dot{1}}} + q^{\dot{A}} \frac{\partial}{\partial p^{\dot{A}}} \Big)$ &  $- \delta_{0} p^{\dot{2}} \frac{\partial}{\partial p^{\dot{2}}} + 2\chi_{0} p^{\dot{A}} \frac{\partial}{\partial p^{\dot{A}}}$
& & 
\\ \hline
$\pounds_{K} \Theta=2\chi_{0} \Theta$ & $\pounds_{K} \Theta= (6\chi_{0} - 2\delta_{0})\Theta$ &
$\pounds_{K} \Theta=6\chi_{0} \Theta$ & $\pounds_{K} \Theta=6 \chi_{0} \Theta$
\\ \hline
$\Theta=q^{\dot{1}} \, T(q^{\dot{2}},x,y)$ & 
$\Theta=(q^{\dot{1}})^{\frac{6\chi_{0}-2\delta_{0}}{\delta_{0}}} \, T(q^{\dot{2}},x,y)$ &
$\Theta=e^{6 \chi_{0}q^{\dot{1}}} \, T(q^{\dot{2}},x,y)$ & $\Theta = (p^{\dot{2}})^{3} \, T(q^{\dot{M}}, x)$
\\
$x:=\frac{p^{\dot{1}}}{q^{\dot{1}}}-\ln q^{\dot{1}}$ & 
$x:= p^{\dot{1}} \, (q^{\dot{1}})^{-\frac{2\chi_{0}}{\delta_{0}}}$ &
$x:= p^{\dot{1}} \, e^{-2\chi_{0}q^{\dot{1}}}$ & $x:=\frac{p^{\dot{1}}}{p^{\dot{2}}}$
\\
$y:=\frac{p^{\dot{2}}}{q^{\dot{2}}}-\ln q^{\dot{1}}$ & 
$y:= p^{\dot{2}} \, (q^{\dot{1}})^{-\frac{2\chi_{0}-\delta_{0}}{\delta_{0}}}$ &
$y:= p^{\dot{2}} \, e^{-2\chi_{0}q^{\dot{1}}}$ &
\\ \hline
\end{tabular}
\newline
\newline
After some calculations we can find the reduced forms of the heavenly equation for above types to be
\newline
\newline
\textbf{Type $\mathcal{H}$HKI}
\begin{equation}
\label{niebo_homotetyczne_1}
T_{xx}T_{yy} - T_{xy}^{2} + q^{\dot{2}} \, \big( T_{y} - T_{yy} - (1+x-y) \, T_{xy} \big) - (q^{\dot{2}})^{2} \, T_{x q^{\dot{2}}}=0 
\end{equation}
\textbf{Type $\mathcal{H}$HKII}
\begin{equation}
\label{niebo_homotetyczne_2}
T_{xx}T_{yy} - T_{xy}^{2} + \frac{4\chi_{0}-\delta_{0}}{\delta_{0}} \, T_{y} - \frac{2\chi_{0}}{\delta_{0}} \, xT_{xy} - \frac{2\chi_{0}-\delta_{0}}{\delta_{0}} \, yT_{yy} - T_{x q^{\dot{2}}} = 0
\end{equation}
\textbf{Type $\mathcal{H}$HKIIIa}
\begin{equation}
\label{niebo_homotetyczne_3}
T_{xx}T_{yy} - T_{xy}^{2} + 2\chi_{0} \big( 2 T_{y} - x \, T_{xy} - y \, T_{yy} \big) - T_{x q^{\dot{2}}}=0
\end{equation}
\textbf{Type $\mathcal{H}$HKIIIb}
\begin{equation}
\label{niebo_homotetyczne_4}
6 T T_{xx} -4  T_{x}^{2} - T_{x q^{\dot{2}}} - 3 T_{q^{\dot{1}}} -x \, T_{x q^{\dot{1}}}=0
\end{equation}
We are going to investigate (\ref{niebo_homotetyczne_1}) - (\ref{niebo_homotetyczne_4}) soon.

\setlength\arraycolsep{2pt}
\setcounter{equation}{0}

\section{Appendix.}
\label{redukcja_rownan_killinga}

\subsection{Explicit form of the conformal Killing equations and their integrability conditions in nonexpanding hyperheavenly spaces with $\Lambda$.}

Equations (\ref{rozklad_rownania_Killinga}) - (\ref{definicja_chi}) together with their integrability conditions (\ref{integrability_c_L}) - (\ref{integrability_c_R}) form our problem to be solved. Using the formula for the spinorial covariant derivative
\begin{eqnarray}
\nabla_{M \dot{N}} \Psi^{A \dot{B}}_{C \dot{D}} &=& \partial_{M \dot{N}} \Psi^{A \dot{B}}_{C \dot{D}} 
+ \mathbf{\Gamma}^{A}_{\ SM \dot{N}} \, \Psi^{S \dot{B}}_{C \dot{D}} 
- \mathbf{\Gamma}^{S}_{\ CM \dot{N}} \, \Psi^{A \dot{B}}_{S \dot{D}}
\\ \nonumber
&&+ \mathbf{\Gamma}^{\dot{B}}_{\ \dot{S}M \dot{N}} \, \Psi^{A \dot{S}}_{C \dot{D}}
- \mathbf{\Gamma}^{\dot{S}}_{\ \dot{D}M \dot{N}} \, \Psi^{A \dot{B}}_{C \dot{S}}
\end{eqnarray}
then the decomposition (\ref{rozklad_koneksji}) and the formulae (\ref{rozklad_operatora}) and (\ref{rozklad_wektora_Killinga}), after some work one obtains the system of equations: 
\newline
Conformal Killing equations
\begin{subequations}
\begin{eqnarray}
\label{rowwnania_Killinga}
E_{11}^{\ \ \dot{A}\dot{B}} &\equiv& 2  \, \partial^{(\dot{A}}   k^{\dot{B})}  = 0
\\ 
E_{12}^{\ \ \dot{A}\dot{B}} &\equiv& \partial^{(\dot{A}} \big( h^{\dot{B})} - k_{\dot{S}} \, Q^{\dot{B})\dot{S}} \big) + \frac{\partial k^{(\dot{A}}}{\partial q_{\dot{B})}} + Q^{\dot{A}\dot{B}} \, \partial^{\dot{S}} k_{\dot{S}} = 0
\\ 
E_{22}^{\ \ \dot{A}\dot{B}} &\equiv& 2 \, \eth^{(\dot{A}}h^{\dot{B})} + 2 \, \eth_{\dot{S}}Q^{\dot{S}(\dot{A}} k^{\dot{B})} - 2 \, h^{\dot{S}} \partial_{\dot{S}} Q^{\dot{A}\dot{B}} = 0
\\
\label{the_E_equation}
\frac{1}{2} E &\equiv& 4\chi -\eth^{\dot{N}} k_{\dot{N}} + \partial^{\dot{N}} h_{\dot{N}} -  k_{\dot{S}} \, \partial_{\dot{A}}  Q^{\dot{A}\dot{S}} = 0
\ \ \ \ \ \ \ \ \ \ \ \ \ \ \ \ \ \ \ \ \ \ \ \ \  \ \ \ \ \ \ 
\end{eqnarray}
\end{subequations}
then
\begin{eqnarray}
\label{definicje_spinorow_LAB}
l_{11} &=& \partial^{\dot{N}} k_{\dot{N}}
\\ \nonumber
2 \, l_{12} &=& \partial_{\dot{N}} h^{\dot{N}} + \eth_{\dot{N}} k^{\dot{N}} + k_{\dot{N}} \, \partial_{\dot{S}} Q^{\dot{S}\dot{N}}
\\ \nonumber
l_{22} &=& \eth_{\dot{N}} h^{\dot{N}} + k_{\dot{N}} \, \eth_{\dot{S}} Q^{\dot{S}\dot{N}} 
\ \ \ \ \ \ \ \ \ \ \ \ \ \ \ \ \ \ \ \ \ \ \ \ \ \ \ \ \ \ \ \ \ \ \ \ \ \ \ \ \ \ \ \ \ \ \ \ \ \ 
\end{eqnarray}
Integrability conditions $L_{RST}^{\ \ \ \ \; \dot{A}}$
\begin{eqnarray}
-\frac{1}{\sqrt{2}} \, L_{111}^{\ \ \ \; \dot{A}} &\equiv&   \partial^{\dot{A}}   l_{11}  = 0
\\ \nonumber
-\frac{1}{\sqrt{2}} \, L_{112}^{\ \ \ \; \dot{A}} &\equiv&  \partial^{\dot{A}} (l_{12}+\chi)  +k^{\dot{A}} \Lambda   = 0
\\ \nonumber
-\frac{1}{\sqrt{2}} \, L_{122}^{\ \ \ \; \dot{A}} &\equiv&  \partial^{\dot{A}} l_{22}  - C^{(2)} k^{\dot{A}} +2 \, \eth^{\dot{A}} \chi =0
\\ \nonumber
-\frac{1}{\sqrt{2}} \, L_{211}^{\ \ \ \; \dot{A}} &\equiv&  \eth^{\dot{A}} l_{11} +  l_{11} \, \partial_{\dot{S}} Q^{\dot{S}\dot{A}}  -2 \, \partial^{\dot{A}} \chi  = 0
\\ \nonumber
-\frac{1}{\sqrt{2}} \, L_{212}^{\ \ \ \; \dot{A}} &\equiv&  \eth^{\dot{A}} (l_{12} - \chi) + l_{11} \, \eth_{\dot{S}}Q^{\dot{S}\dot{A}} - h^{\dot{A}} \Lambda  - k^{\dot{A}} C^{(2)}=0
\\ \nonumber
-\frac{1}{\sqrt{2}} \, L_{222}^{\ \ \ \; \dot{A}} &\equiv&  \eth^{\dot{A}} l_{22} - l_{22} \, 
\partial_{\dot{S}} Q^{\dot{S}\dot{A}}  + 2  \, l_{12} \, \eth_{\dot{S}} Q^{\dot{S}\dot{A}}
+ C^{(2)} h^{\dot{A}} - C^{(1)} k^{\dot{A}} =0
\end{eqnarray}
Integrability conditions $M_{ABCD}$
\begin{eqnarray}
\label{wwarunki_MABCD}
M_{1111} &\equiv& 0
\\ \nonumber
M_{1112} &\equiv& -\frac{3}{2}  C^{(3)} \, \partial^{\dot{N}}  k_{\dot{N}} = 0
\\ \nonumber
M_{1122} &\equiv& h_{\dot{N}} \, \partial^{\dot{N}} C^{(3)} - k_{\dot{N}} \, \eth^{\dot{N}} C^{(3)} - 2\chi \, C^{(3)} - \, C^{(2)} \, \partial^{\dot{N}}  k_{\dot{N}} = 0
\\ \nonumber
M_{1222} &\equiv& h_{\dot{N}} \, \partial^{\dot{N}} C^{(2)} - k_{\dot{N}} \, \eth^{\dot{N}} C^{(2)} +C^{(2)} \, \big[ -2\chi -l_{12} +  k_{\dot{N}} \, \partial_{\dot{S}} Q^{\dot{S}\dot{N}}  \big]
\\ \nonumber
&&+\frac{3}{2} C^{(3)} \, \big[l_{22} - 2 k_{\dot{N}} \, \eth_{\dot{S}}Q^{\dot{S}\dot{N}} \big] - \frac{1}{2}  C^{(1)} \, \partial^{\dot{N}} k_{\dot{N}} = 0
\\ \nonumber
M_{2222} &\equiv& h_{\dot{N}} \, \partial^{\dot{N}} C^{(1)} - k_{\dot{N}} \, \eth^{\dot{N}} C^{(1)}
-2C^{(2)} \big[ 2 \, k_{\dot{N}} \, \eth_{\dot{S}} Q^{\dot{S}\dot{N}} - l_{22} \big]
\\ \nonumber
&& +2C^{(1)} \, \big[  k_{\dot{N}} \, \partial_{\dot{S}}  Q^{\dot{N}\dot{S}}  - l_{12} - \chi \big]
\end{eqnarray}
Integrability conditions $N_{AB}^{\ \ \ \dot{A}\dot{B}}$
\begin{eqnarray}
N_{11}^{\ \ \dot{A}\dot{B}} &\equiv& 2  \, \partial^{(\dot{A}}  \partial^{\dot{B})} \chi = 0
\\ \nonumber
N_{12}^{\ \ \dot{A}\dot{B}} &\equiv& N_{21}^{\ \ \dot{B}\dot{A}} 
\\ \nonumber
N_{21}^{\ \ \dot{A}\dot{B}} &\equiv& 2 \, \frac{\partial }{\partial q_{\dot{A}}} \partial^{\dot{B}} \chi 
  + \frac{2}{3} \Lambda \chi  \in^{\dot{A}\dot{B}}
+ 2 \, \partial_{\dot{S}} \chi \cdot \partial^{\dot{B}} Q^{\dot{S}\dot{A}}
+ 2\, Q^{\dot{A}\dot{S}} \, \partial_{\dot{S}} \partial^{\dot{B}} \chi =0
\\ \nonumber
N_{22}^{\ \ \dot{A}\dot{B}} &\equiv& 2 \, \eth^{\dot{A}} \eth^{\dot{B}} \chi +2 \, \partial^{\dot{B}} \chi \cdot \eth_{\dot{S}} Q^{\dot{S}\dot{A}} - 2 \, \eth^{\dot{S}} \chi \cdot \partial_{\dot{S}} Q^{\dot{A}\dot{B}} =0
\end{eqnarray}
Integrability conditions $R_{ABC}^{\ \ \ \ \ \dot{A}}$
\begin{eqnarray}
R_{111}^{\ \ \ \dot{A}} &\equiv& 0
\\ \nonumber
R_{112}^{\ \ \ \dot{A}} &\equiv& \frac{\sqrt{2}}{2} \, C^{(3)} \, \partial^{\dot{A}} \chi =0
\\ \nonumber
R_{122}^{\ \ \ \dot{A}} &\equiv& \frac{\sqrt{2}}{2} \bigg( C^{(2)} \, \partial^{\dot{A}} \chi -C^{(3)} \, \eth^{\dot{A}} \chi \bigg) =0
\\ \nonumber
R_{222}^{\ \ \ \dot{A}} &\equiv& \frac{\sqrt{2}}{2} \bigg( C^{(1)} \, \partial^{\dot{A}} \chi -C^{(2)}  \, \eth^{\dot{A}} \chi \bigg) =0
\ \ \ \ \ \ \ \ \ \ \ \ \ \ \ \ \ \ \ \ \ \ \ \ \ \ \ \ \ \ \ \ \ \ \ \ \ \ \ \ \ \ \ \ \ \ \ \ \ \ \ \
\end{eqnarray}

\subsection{Reduction of the Killing conformal equations to one master equation.}

In the case of hyperheavenly spaces with $\Lambda=0$ such a reduction was done in \cite{biblio_27}. In our work \cite{biblio_51} the existence of nonzero cosmological constant has been considered and a nonconstant, conformal factor $\chi$ has appeared. However, considerations in \cite{biblio_51} are divided into three subcases. Here we present some different approach in which Killing equations in hyperheavenly spaces of the types $[\textrm{II,D,III,N}] \otimes [\textrm{any}]$ are being reduced simultaneously. This way is much more elegant then that presented in \cite{biblio_51} . The only assumption is that $|C^{(3)}|+ |C^{(2)}|+ |C^{(1)}| \ne 0$, i.e., heavenly spaces are excluded from our considerations. That case was solved by Plebański and Finley in \cite{biblio_14}  and generalized in \cite{biblio_51}. Its spinorial version together with classification of the Killing vectors was presented in \cite{biblio_16}. 

With $|C^{(3)}|+ |C^{(2)}|+ |C^{(1)}| \ne 0$ assumed, from the conditions $R_{ABC}^{\ \ \ \ \ \dot{A}}$ it follows that $\partial^{\dot{A}} \chi =0$, so conformal factor can be at most a function of $q^{\dot{M}}$ only, $\chi=\chi(q^{\dot{M}})$. From the $M_{ABCD}$ equations it follows that $\partial^{\dot{N}} k_{\dot{N}}=0$, from $E_{11}^{\ \ \dot{A}\dot{B}}$ we find that $\partial^{(\dot{A}} k^{\dot{B})}=0$ so $k^{\dot{A}}=\delta^{\dot{A}} (q^{\dot{M}})$, and, finally $l_{11}=0$. Putting this into $N_{21}^{\ \ \dot{A}\dot{B}}$ and $R_{222}^{\ \ \ \dot{A}} $ we get
\begin{equation}
\Lambda \chi=0 \ \ \ , \ \ \ C^{(2)} \frac{\partial \chi}{\partial q^{\dot{A}}}=0
\end{equation}

Denoting
\begin{equation}
\label{definicja_XB}
X^{\dot{B}} := h^{\dot{B}} - \delta_{\dot{S}} Q^{\dot{S}\dot{B}} + \frac{\partial \delta^{\dot{B}}}{\partial q_{\dot{S}}} p_{\dot{S}}
\end{equation}
one can rewrite the equation $E_{12}^{\ \ \dot{A}\dot{B}}$ in the form $\partial^{(\dot{A}} X^{\dot{B})}=0$. Hence, $X^{\dot{B}} = X  p^{\dot{B}} - \epsilon^{\dot{B}}$ with $X=X(q^{\dot{M}}) $ and $\epsilon^{\dot{B}}=\epsilon^{\dot{B}}(q^{\dot{M}})$. Inserting $h^{\dot{B}} - \delta_{\dot{S}} Q^{\dot{S}\dot{B}}$ taken from (\ref{definicja_XB}) into the (\ref{the_E_equation}) one obtains $X$
\begin{equation}
X=\frac{\partial \delta^{\dot{N}}}{\partial q^{\dot{N}}} - 2 \chi
\end{equation}
and, finally, $h^{\dot{B}}$ reads
\begin{equation}
h^{\dot{A}} - \delta_{\dot{S}} Q^{\dot{S}\dot{A}} = R^{\dot{A}}
\ \ \ \ \ , \ \ \ \ \ 
R^{\dot{A}} := - 2\chi \, p^{\dot{A}} - \frac{\partial \delta^{\dot{N}}}{\partial q_{\dot{A}}} \, p_{\dot{N}} - \epsilon^{\dot{A}}
\end{equation}
The next step is to calculate $l_{12}$ from the definition (\ref{definicje_spinorow_LAB})
\begin{equation}
l_{12} = \delta_{\dot{N}} (F^{\dot{N}} + \Lambda p^{\dot{N}}) - 2\chi + \frac{\partial \delta^{\dot{N}}}{\partial q^{\dot{N}}}
\end{equation}
With this form of $l_{12}$ we easily find, that $L_{112}^{\ \ \ \; \dot{A}}$ is automatically satisfied but $L_{212}^{\ \ \ \; \dot{A}}$ gives the integrability condition
\begin{equation}
-\Lambda \epsilon^{\dot{A}} = \delta^{\dot{A}} \frac{\partial F^{\dot{M}}}{\partial q^{\dot{M}}} + \frac{\partial}{\partial q_{\dot{A}}} \Big( \delta_{\dot{N}} F^{\dot{N}} + \frac{\partial \delta^{\dot{N}}}{\partial q^{\dot{N}}} - 3 \chi \Big)
\end{equation}
Analogously, investigating the definition of $l_{22}$ we get
\begin{equation}
\label{definicja_el_22}
l_{22} = -C^{(2)} \, \delta_{\dot{N}} p^{\dot{N}} -2 \, \frac{\partial \chi}{\partial q^{\dot{N}}} \, p^{\dot{N}} + 2 \delta_{\dot{N}} N^{\dot{N}} - \frac{\partial \epsilon^{\dot{N}}}{\partial q^{\dot{N}}}
\end{equation}
Inserting this $l_{22}$ into the $L_{122}^{\ \ \ \; \dot{A}}$ we get an identity, but $L_{222}^{\ \ \ \; \dot{A}}$ leads to the equation
\begin{equation}
\label{calka_N22AB}
\frac{\partial \chi}{\partial q^{\dot{N}}} \Theta_{p_{\dot{N}}p_{\dot{A}}} - p_{\dot{N}} \bigg( \frac{\partial^{2} \chi}{\partial q_{\dot{N}} \partial q_{\dot{A}}} - \frac{2}{3} \frac{\partial \chi}{\partial q_{(\dot{N}}} F^{\dot{A})} \bigg) + \frac{1}{2} G^{\dot{A}} = 0
\end{equation}
with $G^{\dot{A}}$ given by (\ref{definicja_GA}). It is easy to note that (\ref{calka_N22AB}) is just a first integral of $N_{22}^{\ \ \dot{A}\dot{B}}$ and in the case of conformal Killing symmetries it gives some strong constraint on the key function $\Theta$.

Two useful equations follow from the integrability condition $M_{2222}$ which appears to be linear in $p_{\dot{A}}$. That equations read
\begin{eqnarray}
&&\frac{\partial}{\partial q^{\dot{N}}} (\delta^{\dot{N}}Z^{\dot{A}} ) + \frac{\partial \delta_{\dot{N}}}{\partial q_{\dot{A}}} Z^{\dot{N}} =0 \ \ \ \leftrightarrow \ \ \ \frac{\partial}{\partial q_{\dot{A}}} \big( \delta_{\dot{N}} Z^{\dot{N}} \big) + Z^{\dot{A}} \frac{\partial \delta^{\dot{N}}}{\partial q^{\dot{N}}} + \delta^{\dot{A}} \frac{\partial Z^{\dot{N}}}{\partial q^{\dot{N}}}=0 
\\
\label{warunek_na_pochodna_GA}
&& \frac{\partial G^{\dot{A}}}{\partial q^{\dot{A}}} + F_{\dot{A}}G^{\dot{A}} - 2 N^{\dot{A}} \frac{\partial \chi}{\partial q^{\dot{A}}}=0 \ \ \ \ \ \
\end{eqnarray}
with $Z^{\dot{A}}$ defined by (\ref{definicja_Z}). 

Eq. (\ref{warunek_na_pochodna_GA}) is especially interesting. It plays role only in the case of conformal symmetries (observe that if conformal factor is constant then from (\ref{calka_N22AB}) it follows that $G^{\dot{A}} =0$). Using hyperheavenly equation, (\ref{calka_N22AB}) and the first derivative of (\ref{calka_N22AB}) one can show that it is identically satisfied. However, (\ref{warunek_na_pochodna_GA}) does not contain the key function $\Theta$ and can be treated as a compatibility condition for hyperheavenly equation and (\ref{calka_N22AB}).

Next step is to integrate the last triplet of the Killing equations, i.e., the $E_{22}^{\ \ \dot{A}\dot{B}} $ equation. After some obvious cancelations, one can get 
\begin{equation}
\frac{1}{2}  E_{22}^{\ \ \dot{A}\dot{B}} \equiv \frac{\partial R^{(\dot{A}}}{\partial q_{\dot{B})}} + Q^{\dot{S}(\dot{A}} \, \partial_{\dot{S}}R^{\dot{B})} - R^{\dot{S}} \, \partial_{\dot{S}} Q^{\dot{A}\dot{B}} + \frac{\partial \delta_{\dot{S}}}{\partial q_{(\dot{A}}} \, Q^{\dot{B})\dot{S}} + \delta^{\dot{S}} \, \frac{\partial Q^{\dot{A}\dot{B}}}{\partial q^{\dot{S}}} = 0
\end{equation}
Inserting $Q^{\dot{A}\dot{B}}$ given by (\ref{expanding_QAB}) after some algebraic work one can bring $E_{22}^{\ \ \dot{A}\dot{B}}$ equation to the form
\begin{equation}
\label{pierwsza_redukcja_ostatniego_tripletu}
\frac{1}{2} E_{22}^{\ \ \dot{A}\dot{B}} \equiv \partial^{(\dot{A}} \Sigma^{\dot{B})} = 0 
\end{equation}
where
\begin{eqnarray}
\label{definicja_sigmaA}
\Sigma^{\dot{A}} &:=& \delta^{\dot{S}} \, \frac{\partial \Omega^{\dot{A}}}{\partial q^{\dot{S}}} - R^{\dot{S}} \, \partial_{\dot{S}} \Omega^{\dot{A}} - 4 \chi \, \Omega^{\dot{A}} + \frac{\partial \delta_{\dot{S}}}{\partial q_{\dot{A}}} \, \Omega^{\dot{S}} + \frac{\partial \delta^{\dot{S}}}{\partial q^{\dot{S}}} \, \Omega^{\dot{A}}
- \frac{\partial \epsilon^{\dot{A}}}{\partial q^{\dot{N}}} \, p^{\dot{N}}
\\ \nonumber
&& -\frac{2}{3} \, p^{\dot{A}}p_{\dot{N}} \bigg( \delta^{\dot{N}} \frac{\partial F^{\dot{M}}}{\partial q^{\dot{M}}} + \frac{\partial}{\partial q_{\dot{N}}} (\delta_{\dot{M}} F^{\dot{M}} ) \bigg) 
-\frac{1}{3} \, p_{\dot{N}} p_{\dot{M}} \bigg( \frac{\partial^{2} \delta^{\dot{N}}}{\partial q_{\dot{A}} \partial q_{\dot{M}}} + \frac{1}{2} \frac{\partial^{2} \delta^{\dot{A}}}{\partial q_{\dot{N}} \partial q_{\dot{M}}} \bigg)
\ \ \ \ \ \ \ 
\end{eqnarray} 
From (\ref{pierwsza_redukcja_ostatniego_tripletu}) it follows that
\begin{equation}
\label{wnioski_z_redukcji_trzeciego_tripletu}\Sigma^{\dot{A}} = \Sigma \, p^{\dot{A}} + \zeta^{\dot{A}} 
\ \ \ , \ \ \ \Sigma=\Sigma(q^{\dot{M}})
\ \ \ , \ \ \ \zeta^{\dot{A}} = \zeta^{\dot{A}}(q^{\dot{M}})
\end{equation}
The next step is to use (\ref{postac_Omegi}) and put it into (\ref{definicja_sigmaA}). 
Most of the factors in (\ref{definicja_sigmaA}) can be rearranged into the form $\partial^{\dot{A}} (...)$. Using (\ref{wnioski_z_redukcji_trzeciego_tripletu}) and (\ref{definicja_sigmaA}) we get
\begin{eqnarray}
\label{drugi_krok_redukcji}
\Sigma \, p^{\dot{A}} + \zeta^{\dot{A}} &=& \partial^{\dot{A}} \bigg( - \pounds_{K} \Theta + 2 \Theta \Big(3 \chi - \frac{\partial \delta^{\dot{N}}}{\partial q^{\dot{N}}} \Big) - \frac{1}{6} \frac{\partial^{2} \delta^{\dot{N}}}{\partial q_{\dot{M}} \partial q_{\dot{S}}} \, p_{\dot{N}} p_{\dot{M}} p_{\dot{S}} \bigg) 
 \ \ \ \ \ \ 
\\ \nonumber
&&- \frac{2}{3} F_{\dot{S}} \, (p^{\dot{S}} \epsilon^{\dot{A}} + p^{\dot{A}} \epsilon^{\dot{S}} ) - \frac{\partial \epsilon^{\dot{A}}}{\partial q^{\dot{N}}} \, p^{\dot{N}}
\end{eqnarray}
Calculating $\partial_{\dot{A}}(\ref{drugi_krok_redukcji})$ one can find $\Sigma$
\begin{equation}
\label{definicja_Sigma}
\Sigma=F^{\dot{S}} \epsilon_{\dot{S}} - \frac{1}{2} \frac{\partial \epsilon^{\dot{S}}}{\partial q^{\dot{S}}}
\end{equation}
With $\Sigma$ given by (\ref{definicja_Sigma}) we can finalize the process of reduction of the third triplet of the Killing equations and bring (\ref{drugi_krok_redukcji}) into the form
\begin{equation}
\partial^{\dot{A}} \bigg( - \pounds_{K} \Theta + 2 \Theta \Big(3 \chi - \frac{\partial \delta^{\dot{N}}}{\partial q^{\dot{N}}} \Big) + \mathcal{P} \bigg) = 0
\end{equation}
with $\mathcal{P}$ defined by (\ref{definicja_wielomianu_P}). In that way we obtain the master equation (\ref{master_equation}). 

Additionally, one can investigate $\eth_{\dot{A}} \partial^{\dot{A}} ( \textrm{master equation} )$. After very laborious calculations we find that it is simply integral of the (\ref{calka_N22AB}) and it leads to the condition (\ref{integrability_condition_6}).

\section{Concluding remarks}

In this paper we end the theoretical considerations on the Killing symmetries in nonexpanding $\mathcal{HH}$-spaces with $\Lambda$. Results from \cite{biblio_51} and \cite{biblio_27} have been gathered in a compact form. Classification of these symmetries is the original part of the theory of the nonexpanding hyperheavenly spaces. Reduction of the ten Killing equations to one master equation has been done under the  assumption, that $C_{ABCD} \ne 0$. However, the final results can be quickly carry over to the case of the isometric and homothetic symmetries in heavenly spaces ($C_{ABCD}=0$). This allows us to classify homothetic Killing symmetries in heavenly spaces. Only the conformal Killing symmetries in heavenly spaces are outside the scope of this paper but they have been considered in \cite{biblio_51}.

Isometric symmetries in heavenly spaces leads to the five different types of the heavenly spaces. Four of them were completely solved, in the fifth case heavenly equation was reduced to the new distinguished equation called the  Boyer-Finley-Plebański equation \cite{biblio_16}. This case is characterized by $\frac{\partial \delta^{\dot{A}}}{\partial q^{\dot{A}}}$ and in heavenly space with second heavenly equation this spinorial divergence cannot be gauged away. In hyperheavenly spaces the gauge freedom seems to be wider than the one in heavenly spaces. Finally, $\frac{\partial \delta^{\dot{A}}}{\partial q^{\dot{A}}}$ can be always gauged away (compare (\ref{divergencja_delty})). This is the main reason, why there is no hyperheavenly equivalent of the Boyer-Finley-Plebański equation. 

An interesting part of our work consists of presenting the solution of the hyperheavenly space admitting isometric Killing vector of the type IK2b. It appeared to be a complex pp-wave. In this case we have been able to find the Lorentzian real slice in a very easy way. Using the results of \cite{biblio_17}, especially condition $\bar{C}_{ABCD} = C_{\dot{A}\dot{B}\dot{C}\dot{D}}$ and substituting (\ref{podstawienia}) we immediately find the real metric with Lorentzian signature describing the real pp-wave (\ref{metryka_dla_pp_fali_lorentzowskiej}). Important question arises: does the condition $\bar{C}_{ABCD} = C_{\dot{A}\dot{B}\dot{C}\dot{D}}$ can be used in easy way in the more complicated problems, for example, complex metrics of the expanding type $[\textrm{N}] \otimes [\textrm{N}]$ found in \cite{biblio_54}?

Finally, the results presented in this paper can be used to describe the symmetries of Einstein - Walker spaces. All Einstein - Walker spaces have been found in \cite{biblio_32}. It is enough to take the metrics from \cite{biblio_32} and substituting them into the master equation and its integrability conditions. Solutions of these equations give the Einstein - Walker space with additional, Killing symmetry. Can Einstein - Walker spaces with Killing symmetry can be classified according to the type of the Killing vector? We are going to answer this question soon.

This paper is the third part of the work devoted to the Killing symmetries in $\mathcal{HH}$-spaces. In the last part we will deal with the conformal symmetries in heavenly spaces with $\Lambda$. Nonzero $\Lambda$ change the structure of the self-dual (anti-self-dual) heavenly spaces - there is no nonexpanding congruences of anti-self-dual (self-dual) null strings. One must use the expanding hyperheavenly equation with the condition $C_{ABCD}=0$. Does the (conformal) Killing equations can be reduced to one master equation in this case? 
\newline
\newline
\textbf{Acknowledgments.} 
\newline
I am indebted to prof. M. Przanowski for many enlightening discussions and help in many crucial matters.

\end{document}